\documentclass[preprint,lineno]{revtex4-2}

\usepackage{natbib}
\usepackage{color}
\usepackage{verbatim}
\usepackage{graphicx}
\usepackage{psfrag}
\usepackage{bm}
\usepackage{longtable}
\usepackage{array}
\usepackage{amsmath}
\usepackage{amssymb}
\usepackage{cancel}
\usepackage{stackengine}
\begin{document}
\definecolor{brown}{rgb}{0.65, 0.16, 0.16}
\definecolor{green}{rgb}{.05,.5,.06}
\renewcommand{\baselinestretch}{1.1}
\def\labelp#1{\label{#1}}
\def\barF{\bar{F}}
\def\barf{\bar{f}}
\def\Sigmap{\Sigma}
  \def\bec{}
  \def\be{\beta}
 \def\bA{{\bf A}}
 \def\bB{\bm{B}}
 \def\bE{\bm{E}}
 \def\bJ{\bm{J}}
 \def\bme{\hat{\bm{e}}}
 \def\bmep{\hat{\bm{e}}^\prime}
 \def\bOmega{\bm{\Omega}}
  \def\bOmegaa{\bm{\Omega}^\ast}
 \def\bx{\bm{x}}
 \def\bxp{\bm{x}^\prime}
 \def\bxpp{\bm{x}^{\prime \prime}}
 \def\xpp{x^{\prime \prime}}
 \def\xp{x^\prime}
 \def\bxs{\bm{x}^S}
 \def\xs{x^S}
 \def\xp{x^\prime}
 \def\bR{\bm{R}}
 \def\bu{\bm{u}}
 \def\bcdot{\bm{\cdot}}
 \def\ra{r^\ast}
 \def\Ra{R^\ast}
 \def\Omegaa{\Omega^\dag}
 \def\Omegad{\Omega^\ast}
 \def\Omegaaa{\Omega^{\ddag}}
 \def\cF{{\cal F}}
 \def\bep{\beta_p}
 \def\bes{\beta_s}
 \def\bech{\beta_{ch}}
 \def\Sigs{\Sigmap}
 \def\Sigp{\Sigmap}
 \def\calC{{\cal C}}
 \def\calP{{\cal P}}
 \def\bomega{\bm{\omega}}
 \def\barbomega{\bar{\bm{\omega}}}
 \def\bOmega{\bm{\Omega}}
 \def\bR{{\cal \bf R}}
 \def\ck{k}
 \def\Ri{R_i}
 \def\bM{\bm{M}}
 \def\bMa{\bm{M}^\ast}
 \def\Ria{\delta}
 \def\bT{\bm{T}}
 \def\dbT{\delta \hspace{-0.3mm} \bm{T}}
 \def\dT{\delta \hspace{-0.3mm} T}
 \def\ckdag{k^{\ddag}}
 \def\bhOmega{\hat{\bm{\Omega}}}
 \def\bhH{\hat{\bm{H}}}
 \def\hOmega{\hat{\Omega}}
 \def\hH{\hat{H}}
 \def\bfe{{\bf e}}
 \def\bht{\hat{\bm{\omega}}}
 \def\bt{\bm{\omega}}
 \def\homega{\hat{\omega}}
 \def\hOmegaone{\hat{\Omega}^{(1)}}
 \def\hOmegatwo{\hat{\Omega}^{(2)}}
 \def\Omegadzero{\Omega^{\dag(0)}}
 \def\Omegadone{\Omega^{\dag(1)}}
 \def\Omegadtwo{\Omega^{\dag(2)}}
 \def\Mizero{M_I^{(0)}}
 \def\Mrzero{M_R^{(0)}}
 \def\Mip{M_I'}
 \def\Mrp{M_R'}
 \def\tH{\hat{\bm{\omega}} \bcdot \hat{\bm{H}}}
 \def\tO{\hat{\bm{\omega}} \bcdot \hat{\bm{\Omega}}}
 \def\tO{\hat{\Omega}_{\bm{\omega}}}
 \def\HO{\hat{\bm{H}} \bcdot \hat{\bm{\Omega}}}
 \def\HO{\hat{\Omega}_{\bm{H}}}
 \def\bOmegad{\bm{\Omega}^\ast}
 \def\bepsilon{\hat{\bm{\epsilon}}}
 \def\bhomega{\hat{\bm{\omega}}}
 \def\hOmegat{\hat{\Omega}_{\bt}}
 \def\hOmegapar{\hat{\Omega}_\parallel}
 \def\hOmegaperp{\hat{\Omega}_\perp}
 \def\bL{{\bf L}}
 \def\bsigmap{\bm{\sigma}^{(p)}}
  \def\bsigmamp{\bm{\sigma}_{M}^{(p)}}
 \def\sigmap{\sigma^{(p)}}
 \def\btimes{\bm{\times}}
 \def\cG{{\cal G}}
 \def\cD{{\cal D}}
 \def\bcG{\bm{{\cal G}}}
 \def\bcD{\bm{{\cal D}}}
\def\bv{\bm{v}}
\def\rhoe{\varrho_c}
\def\bB{\bm{B}}
\def\bF{\bm{F}}
\def\bFh{\bm{F}^h}
\def\bTh{\bm{T}^h}
\def\bTm{\bm{T}^m}
\def\hbB{\hat{\bm{B}}}
\def\br{\bm{r}}
\def\bvo{\bm{v}^{(o)}}
\def\bvi{\bm{v}^{(i)}}
\def\bvoa{\bm{v}^{(o)\dag}}
\def\bvia{\bm{v}^{(i)\dag}}
\def\bI{\bm{I}}
\def\bsigmam{\bm{\sigma}_M}
\def\bepar{\hat{\bm{e}}_\parallel}
\def\beperp{\hat{\bm{e}}_\perp}
\def\bH{\bm{H}}
\def\dbH{\delta \! \bm{H}}
\def\Momega{M_{\bomega}}
\def\Momegaa{M_{\bomega}^\ast}
\def\Mpar{M_{\parallel}}
\def\Mperp{M_{\perp}}
\def\Mpara{M_{\parallel}^\ast}
\def\Mperpa{M_{\perp}^\ast}
\def\bsigmap{\bm{\sigma}^{p}}
\def\bsigmaps{\bm{\sigma}_s^{p}}
\def\bsigmapa{\bm{\sigma}_a^{p}}
\def\bsigmape{\bm{\sigma}_e^{p}}
\def\bsigmapi{\bm{\sigma}_i^{p}}
\def\Momega{M_{\bomega}}
\def\Mpar{M_{\parallel}}
\def\Mperp{M_{\perp}}
\def\Sigch{\Sigma_{ch}}
\def\Lambdap{\Lambda_p}
\def\Lambdas{\Lambda_s}
\def\bomegadag{\bm{\omega}^\dag}
\def\bnabla{\bm{\nabla}}
\def\tbH{\tilde{\bm{H}}}
\def\tbE{\tilde{\bm{E}}}
\def\tH{\tilde{H}}
\def\tE{\tilde{E}}
\def\bA{\bm{A}}
\def\tbA{\tilde{\bm{A}}}
\def\tbB{\tilde{\bm{B}}}
\def\tbBpar{\tilde{\bm{B}}^{{\mbox{\tiny $\parallel$}}}}
\def\tbBperp{\tilde{\bm{B}}^{{\mbox{\tiny $\perp$}}}}
\def\bn{\bm{n}}
\def\bns{\bm{n}^S}
\def\ns{n^{S}}
\def\hbx{\hat{\bm{x}}}
\def\hx{\hat{x}}
\def\bPhi{\bm{\Phi}}
\def\hH{\hat{H}}
\def\hbH{\hat{\bm{H}}}
\def\tA{\tilde{A}}
\def\tB{\tilde{B}}
\def\tAO{\tilde{A}_O}
\def\tAI{A_I}
\def\tAR{A_R}
\def\tB{\tilde{B}}
\def\tBR{B_R}
\def\tBI{B_I}
\def\tBO{\tilde{B}_O}
\def\tC{\tilde{C}}
\def\chiR{\chi_R}
\def\chiI{\chi_I}
\def\chipR{\chi_R^{\prime}}
\def\chipI{\chi_I^{\prime}}
\def\total{\mbox{d}}
\def\tbe{\tilde{\beta}}
\def\tbea{\tilde{\beta}^\ast}
\def\tbeasq{\tilde{\beta}^{\ast 2}}
\def\etai{\eta_i}
\def\etao{\eta_o}
\def\psia{\psi^\dag}
\def\ra{r^\dag}
\def\rasq{r^{\dag 2}}
\def\racu{r^{\dag 3}}
\def\calI{{\cal I}}
\def\psii{\psi^{(i)}}
\def\psiia{\psi^{(i)\dag}}
\def\psioa{\psi^{(o)\dag}}
\def\psio{\psi^{(o)}}
\def\psiio{\psi^{(i/o)}}
\def\psiioa{\psi^{(i/o}\dag}
\def\va{v^\dag}
\def\bva{\bm{v}^\dag}
\def\bv{\bm{v}}
\def\bff{\bm{f}}
\def\barbff{\bar{\bm{f}}}
\def\bffp{\bm{f}^\prime}
\def\bffpp{\bm{f}^{\prime \prime}}
\def\bG{\bar{G}}
\def\barbv{\bar{\bm{v}}}
\def\barv{\bar{v}}
\def\barp{\bar{p}}
\def\barpsi{\bar{\psi}}
\def\barpsii{\bar{\psi}^{(i)}}
\def\barpsio{\bar{\psi}^{(o)}}
\def\barpsiio{\bar{\psi}^{(i/o}}
\def\barpsiioa{\bar{\psi}^{(i/o)\dag}}
\def\barpsiia{\bar{\psi}^{(i)\dag}}
\def\barpsioa{\bar{\psi}^{(o)\dag}}
\def\barva{\bar{v}^\dag}
\def\barg{\bar{g}}
\def\gp{g^\prime}
\def\gpp{g^{\prime \prime}}
\def\csch{\mbox{csch}}
\def\bHR{\bm{H}_R}
\def\bHI{\bm{H}_I}
\def\bER{\bm{E}_R}
\def\bEI{\bm{E}_I}
\def\ER{E_R}
\def\EI{E_I}
\def\HRone{H_R^{(1)}}
\def\HRtwo{H_R^{(2)}}
\def\HIone{H_I^{(1)}}
\def\HItwo{H_I^{(2)}}
\def\frp{f_r^{\prime}}
\def\frpp{f_r^{\prime \prime}}
\def\fthp{f_{\theta}^{\prime}}
\def\fthpp{f_{\theta}^{\prime \prime}}
\def\tbffp{\tilde{\bm{f}}^\prime}
\def\tbff{\tilde{\bm{f}}}
\def\tbff{\tilde{\bm{f}}}
\def\tfr{\tilde{f}_r}
\def\tfth{\tilde{f}_{\theta}}
\def\tFr{\tilde{F}_r}
\def\tFth{\tilde{F}_{\theta}}
\def\tbv{\tilde{\bm{v}}}
\def\tv{\tilde{v}}
\def\tvo{\tilde{v}^{(o)}}
\def\tvi{\tilde{v}^{(i)}}
\def\tvia{\tilde{v}^{(i)\dag}}
\def\etao{\eta_o}
\def\tpsii{\tilde{\psi}^{(i)}}
\def\tpsio{\tilde{\psi}^{(o)}}
\def\tpsiia{\tilde{\psi}^{(i)\dag}}
\def\tpsioa{\tilde{\psi}^{(o)\dag}}
\def\tp{\tilde{p}}
\def\tP{\tilde{P}}
\def\tpi{\tilde{p}^{(i)}}
\def\tPi{\tilde{P}^{(i)}}
\def\tpo{\tilde{p}^{(o)}}
\def\tPo{\tilde{P}^{(o)}}
\def\Re{\mbox{Re}_\omega}
\def\Reo{\mbox{Re}_\omega^{(o)}}
\def\tg{\tilde{g}}
\def\tgo{\tilde{g}^{(o)}}
\def\tgi{\tilde{g}^{(i)}}
\def\bargi{\bar{g}^{(i)}}
\def\bargo{\bar{g}^{(o)}}
\def\barfr{\bar{f}_r}
\def\tal{\tilde{\alpha}}
\def\talo{\tilde{\alpha}_o}
\def\barbvo{\bar{\bm{v}}^{(o)}}
\def\barbvi{\bar{\bm{v}}^{(i)}}
\def\barbvoa{\bar{\bm{v}}^{(o)\dag}}
\def\barbvia{\bar{\bm{v}}^{(i)\dag}}
\def\tbvo{\tilde{\bm{v}}^{(o)}}
\def\tbvi{\tilde{\bm{v}}^{(i)}}
\def\tbvoa{\tilde{\bm{v}}^{(o)\dag}}
\def\tbvia{\tilde{\bm{v}}^{(i)\dag}}
\def\bsigma{\bm{\sigma}}
\def\barsigmap{\bar{\sigma}^{(p)}}
\def\tsigmap{\tilde{\sigma}^{(p)}}
\def\barsigma{\bar{\sigma}}
\def\tsigma{\tilde{\sigma}}
\def\tsigmai{\tilde{\sigma}^{(i)}}
\def\tsigmao{\tilde{\sigma}^{(o)}}
\def\tV{\tilde{V}}
\def\tSigma{\tilde{\Sigma}}
\def\tSigmai{\tilde{\Sigma}^{(i)}}
\def\tSigmao{\tilde{\Sigma}^{(o)}}
\def\scvar{\xi}
\def\tgig{\tgi_g}
\def\tgip{\tgi_p}
\def\bargig{\bargi_g}
\def\bargip{\bargi_p}
\def\tgig{\tgi_g}
\def\tgip{\tgi_p}
\def\barPsi{\bar{\Psi}}
\def\tF{\tilde{F}}
\def\barvr{\barv_r}
\def\barvth{\barv_\theta}
\def\tbJ{\tilde{\bm{J}}}
\def\barbsigma{\bar{\bm{\sigma}}}
\def\tbsigma{\tilde{\bm{\sigma}}}
\def\barV{\bar{V}}
\def\barSigma{\bar{\Sigma}}
\def\Ca{\mbox{Ca}}
\def\barh{\bar{h}}
\def\th{\tilde{h}}
\def\bzeta{\bm{\zeta}}
\def\zetazero{\tilde{\zeta}^{(0)}}
\def\zetazerop{\tilde{\zeta}_0^{\prime}}
\def\zetaone{\tilde{\zeta}^{(1)}}
\def\zetatwo{\tilde{\zeta}^{(2)}}
\def\zetathree{\tilde{\zeta}^{(3)}}
\def\bzetaone{\tilde{\bzeta}^{(1)}}
\def\bzetatwo{\tilde{\bzeta}^{(2)}}
\def\bzetathree{\tilde{\bzeta}^{(3)}}
\def\zetazeroa{\zeta_0^{\ast}}
\def\zetaonea{\tilde{\zeta}^{(1)\ast}}
\def\zetatwoa{\tilde{\zeta}^{(2)\ast}}
\def\zetathreea{\tilde{\zeta}^{(3)\ast}}
\def\bzetazeroa{\tilde{\bzeta}^{(0)\ast}}
\def\bzetaonea{\tilde{\bzeta}^{(1)\ast}}
\def\bzetatwoa{\tilde{\bzeta}^{(2)\ast}}
\def\bzetathreea{\tilde{\bzeta}^{(3)\ast}}
\def\bxi{\bm{\xi}}
\def\xizero{\tilde{\xi}^{(0)}}
\def\xizerop{\tilde{\xi}_0^{\prime}}
\def\xione{\tilde{\xi}^{(1)}}
\def\xitwo{\tilde{\xi}^{(2)}}
\def\xithree{\tilde{\xi}^{(3)}}
\def\bxione{\tilde{\bxi}^{(1)}}
\def\bxitwo{\tilde{\bxi}^{(2)}}
\def\bxithree{\tilde{\bxi}^{(3)}}
\def\Ai{A^{(i)}}
\def\Bi{B^{(i)}}
\def\Ci{\tilde{C}_I}
\def\Di{\tilde{D}_I}
\def\Ei{\tilde{E}_I}
\def\Fi{\tilde{F}_I}
\def\Ao{\tilde{A}_{O}}
\def\Bo{\tilde{B}_{O}}
\def\Co{\tilde{C}_{O}}
\def\Copar{\tilde{C}_{O \parallel}}
\def\Coperp{\tilde{C}_{O \perp}}
\def\Cipar{\tilde{C}_{I \parallel}}
\def\Ciperp{\tilde{C}_{I \perp}}
\def\Do{\tilde{D}_{O}}
\def\Dopar{\tilde{D}_{O}^{\parallel}}
\def\Doperp{\tilde{D}_{O}^{\perp}}
\def\Dopara{\tilde{D}_{O}^{\parallel \ast}}
\def\Doperpa{\tilde{D}_{O}^{\perp \ast}}
\def\Doparperp{\tilde{D}_{O}^{\parallel \perp}}
\def\Doperppar{\tilde{D}_{O}^{\perp \parallel}}
\def\Doperpperp{\tilde{D}_{O}^{\perp \perp}}
\def\Eo{E_{(O)}}
\def\Fo{F^{(o)}}
\def\bPhi{\bm{\Phi}}
\def\Phizero{\Phi^{(0)}}
\def\Phione{\Phi^{(1)}}
\def\Phitwo{\Phi^{(2)}}
\def\Phithree{\Phi^{(3)}}
\def\bPhione{\bPhi^{(1)}}
\def\bPhitwo{\bPhi^{(2)}}
\def\bPhithree{\bPhi^{(3)}}
\def\bPhifour{\bPhi^{(4)}}
\def\Phizerox{\Phi_0(\bm{x})}
\def\Phionex{\Phi^{(1)}(\bm{x})}
\def\Phitwox{\Phi^{(2)}(\bm{x})}
\def\Phithreex{\Phi^{(3)}(\bm{x})}
\def\bPhionex{\bPhi^{(1)}(\bm{x})}
\def\bPhitwox{\bPhi^{(2)}(\bm{x})}
\def\bPhithreex{\bPhi^{(3)}(\bm{x})}
\def\Phizeros{\Phi_0(\bxs)}
\def\Phiones{\Phi^{(1)}(\bxs)}
\def\Phitwos{\Phi^{(2)}(\bxs)}
\def\Phithrees{\Phi^{(3)}(\bxs)}
\def\bPhiones{\bPhi^{(1)}(\bxs)}
\def\bPhitwos{\bPhi^{(2)}(\bxs)}
\def\bPhithrees{\bPhi^{(3)}(\bxs)}

\def\bG{\bm{{\cal G}}}
\def\tphi{\tilde{\phi}}
\def\Gzero{{\cal G}}
\def\bGzero{\bm{{\cal G}}}
\def\bHzero{\bm{{\cal H}}}
\def\Hzero{{\cal H}}
\def\bF{\bm{F}}
\def\bddot{\bm{:}}
\def\Ko{\tilde{K}_O}
\def\Lo{\tilde{L}_O}
\def\texte{\mbox{e}}
\def\bo{\hat{\bm{o}}}
\def\ho{\hat{o}}
\def\bHzeropar{\bHzero^{\mbox{\tiny $\parallel$}}}
\def\bHzeroperp{\bHzero^{\mbox{\tiny $\perp$}}}
\def\bGzeroparpar{\bGzero^{\mbox{\tiny $\parallel \parallel$}}}
\def\bGzeroparperp{\bGzero^{\mbox{\tiny $\parallel \perp$}}}
\def\bGzeroperppar{\bGzero^{\mbox{\tiny $\perp \parallel$}}}
\def\bGzeroperpperp{\bGzero^{\mbox{\tiny $\perp \perp$}}}
\def\Hzeropar{\Hzero^{\mbox{\tiny $\parallel$}}}
\def\Hzeroperp{\Hzero^{\mbox{\tiny $\perp$}}}
\def\Gzeroparpar{\Gzero^{\mbox{ \tiny $\parallel \parallel$}}}
\def\Gzeroparperp{\Gzero^{\mbox{\tiny $\parallel \perp$}}}
\def\Gzeroperppar{\Gzero^{\mbox{\tiny $\perp \parallel$}}}
\def\Gzeroperpper{\Gzero^{\mbox{ \tiny $\perp \perp$}}}
\def\bxpar{\bx^{\! \! \mbox{ \tiny $\parallel$}}}
\def\bxperp{\bx^{\! \! \mbox{ \tiny $\perp$}}}
\def\tbHpar{\tbH^{\mbox{\tiny $\parallel$}}}
\def\tbHperp{\tbH^{\mbox{\tiny $\perp$}}}
\def\tHpar{\tH^{\mbox{\tiny $\parallel$}}}
\def\tbHpara{\tbH^{\mbox{\tiny $\parallel \ast$}}}
\def\tbHperpa{\tbH^{\mbox{\tiny $\perp \ast$}}}
\def\tHperpa{\tH^{\mbox{\tiny $\perp$ \ast}}}
\def\tHperpa{\tH^{\mbox{\tiny $\perp$} \ast}}
\def\bfpar{\bar{\bm{f}}^{\mbox{\tiny $\parallel$}}}
\def\bfperp{\bar{\bm{f}}^{\mbox{\tiny $\perp$}}}
\def\Int{\tilde{\mbox{I}}}
\def\tC{\tilde{C}}
\def\tphi{\tilde{\phi}}
\def\bK{\bm{K}}
\def\barbK{\bar{\bm{K}}}
\def\tbK{\tilde{\bm{K}}}
\def\barK{\bar{K}}
\def\barKs{\bar{K}^s}
\def\barKa{\bar{K}^a}
\def\tK{\tilde{K}}
\def\tKs{\tilde{K}^s}
\def\tKa{\tilde{K}^a}
\def\barbKs{\bar{\bm{K}}^s}
\def\tbKa{\bar{\bm{K}}^a}
\def\tKs{\tilde{\bm{K}}^s}
\def\tKa{\tilde{\bm{K}}^a}
\def\barbF{\bar{\bm{F}}}
\def\tbF{\tilde{\bm{F}}}
\def\barbFp{\bar{\bm{F}}^\prime}
\def\barGamma{\bar{\Gamma}}
\def\tGamma{\tilde{\Gamma}}
\def\barGamparone{\bar{\Gamma}_{r1}^{\mbox{\tiny $\parallel$}}}
\def\Gzeroamparone{\tilde{\Gamma}_{r1}^{\mbox{\tiny $\parallel$}}}
\def\barGamperpone{\bar{\Gamma}_{r1}^{\mbox{\tiny $\perp$}}}
\def\Gzeroamperpone{\tilde{\Gamma}_{r1}^{\mbox{\tiny $\perp$}}}
\def\barGampartwo{\bar{\Gamma}_{r2}^{\mbox{\tiny $\parallel$}}}
\def\Gzeroampartwo{\tilde{\Gamma}_{r2}^{\mbox{\tiny $\parallel$}}}
\def\barGamperptwo{\bar{\Gamma}_{r2}^{\mbox{\tiny $\perp$}}}
\def\Gzeroamperptwo{\tilde{\Gamma}_{r2}^{\mbox{\tiny $\perp$}}}
\def\barGamperpthree{\bar{\Gamma}_{r3}^{\mbox{\tiny $\perp$}}}
\def\Gzeroamperpthree{\tilde{\Gamma}_{r3}^{\mbox{\tiny $\perp$}}}
\def\barLambdaone{\bar{\Lambda}_{r1}}
\def\barLambdatwo{\bar{\Lambda}_{r2}}
\def\barLambdathree{\bar{\Lambda}_{r3}}
\def\tLambdaone{\tilde{\Lambda}_{r1}}
\def\tLambdatwo{\tilde{\Lambda}_{r2}}
\def\Qparone{\tilde{Q}^{{\mbox{\tiny $\parallel$}}}_1}
\def\Qpartwo{\tilde{Q}^{{\mbox{\tiny $\parallel$}}}_2}
\def\Qparthree{\tilde{Q}^{{\mbox{\tiny $\parallel$}}}_3}
\def\Qperpone{\tilde{Q}^{{\mbox{\tiny $\perp$}}}_1}
\def\Qperptwo{\tilde{Q}^{{\mbox{\tiny $\perp$}}}_2}
\def\Qperpthree{\tilde{Q}^{{\mbox{\tiny $\perp$}}}_3}
\def\Qperpfour{\tilde{Q}^{{\mbox{\tiny $\perp$}}}_4}
\def\Qperpfive{\tilde{Q}^{{\mbox{\tiny $\perp$}}}_5}
\def\Qperpsix{\tilde{Q}^{{\mbox{\tiny $\perp$}}}_6}
\def\Pone{\tilde{P}_1}
\def\Ptwo{\tilde{P}_2}
\def\Pthree{\tilde{P}_3}
\def\Pfour{\tilde{P}_4}
\def\Ponea{\tilde{P}_1^\ast}
\def\Ptwoa{\tilde{P}_2^\ast}
\def\Pthreea{\tilde{P}_3^\ast}
\def\Pfoura{\tilde{P}_4^\ast}
\def\barT{\bar{T}}
\def\bTh{\bm{T}^h}
\def\barbT{\bar{\bm{T}}}
\def\barbTprime{\bar{\bm{T}}^\prime}
\def\tT{\tilde{T}}
\def\tbT{\tilde{\bm{T}}}
\def\barbTperp{\bar{\bm{T}}^{{\mbox{\tiny $\perp$}}}}
\def\tbTperp{\tilde{\bm{T}}^{{\mbox{\tiny $\perp$}}}}
\def\aint{\int_{z,C}}
\def\barbFpar{\bar{\bm{F}}^{{\mbox{\tiny $\parallel$}}}}
\def\barbFperp{\bar{\bm{F}}^{{\mbox{\tiny $\perp$}}}}
\def\tbFpar{\tilde{\bm{F}}^{{\mbox{\tiny $\parallel$}}}}
\def\tbFperp{\tilde{\bm{F}}^{{\mbox{\tiny $\perp$}}}}
\def\barbFparprime{\bar{\bm{F}}^{{\mbox{\tiny $\parallel$}}'}}
\def\barbFperpprime{\bar{\bm{F}}^{{\mbox{\tiny $\perp$}}'}}
\def\bHzeroprime{\bHzero'}
\def\bGzeroprime{\bGzero'}
\def\xp{x'}
\def\yp{y'}
\def\zp{z'}
\def\bcR{\bm{{\cal R}}}
\def\tchi{\tilde{\chi}}
\def\tbchi{\tilde{\bm{\chi}}}
\def\tchipar{\tilde{\chi}_{\parallel}}
\def\tchiperp{\tilde{\chi}_{\perp}}
\def\tlambda{\tilde{\lambda}}
\def\tchia{\tilde{\chi}^\ast}
\def\tchipara{\tilde{\chi}_{\parallel}^\ast}
\def\tchiperpa{\tilde{\chi}_{\perp}^\ast}
\def\tlambdaa{\tilde{\lambda}^\ast}
\def\rhon{\rho}
\def\barrhon{\bar{\rho}}
\def\barphin{\bar{\upsilon}}
\def\drhon{\delta \! \rho}
\def\dbv{\delta \! \bm{v}}
\def\dbarbF{\delta \! \bar{\bm{F}}}
\def\dv{\delta \! v}
\def\tbHp{\delta \! \tilde{\bm{H}}}
\def\tHp{\delta \! \tilde{H}}
\def\tbGp{\delta \hspace{-0.4mm} \tilde{\bm{G}}}
\def\tGp{\delta \hspace{-0.4mm} \tilde{G}}
\def\tbHpa{\delta \! \tilde{\bm{H}}^{\ast}}
\def\tbGpa{\tbGp^{\ast}}
\def\tHpa{\delta \! \tilde{H}^{\ast}}
\def\tGpa{\delta \! \tilde{G^{\ast}}}
\def\bk{\bm{k}}
\def\bD{\bm{D}}
\def\bDM{\bm{D}^{M}}
\def\DB{D_{B}}
\def\bDMpar{\bm{D}_{\parallel}^{M}}
\def\bDMperp{\bm{D}_{\perp}^{M}}
\def\DMpar{D_{\parallel}^{M}}
\def\DMperp{D_{\perp}^{M}}
\def\kperp{k_{\perp}}
\def\kpar{k_{\parallel}}
\def\hbHzero{\hat{\bm{{\cal H}}}}
\def\dbo{\delta \hspace{-0.4mm} \hat{\bm{o}}}
\def\dho{\delta \hspace{-0.4mm} \hat{o}}
\def\Mupar{\mathrm{M}_{\parallel}}
\def\Muperp{\mathrm{M}_{\perp}}
\def\Rcu{R^3}
\def\Rfo{R^4}
\def\Rfi{R^5}
\def\Rsi{R^6}
\def\RsqL{R^2 L}
\def\RfoLsq{R^4 L^2}

\title[Induced-current magnetophoresis]{Induced-current
magnetophoresis}

%
 \author{V. Kumaran, Department of Chemical Engineering, Indian Institute of Science, Bangalore 560 012, India.}

\begin{abstract}
 When an electrically conducting non-magnetic particle is subjected to a spatially varying and oscillating
 applied magnetic field of amplitude $\bHzero + \bGzero \bcdot \bx$ and frequency $\omega$, an
 oscillating eddy current is induced. The Lorentz force density, the cross
 product of the current density and the magnetic field, consists of a steady component
 and a component with frequency $2 \omega$.
 If there is a spatial variation in the applied field, there is a steady force
 on a sphere of radius $R$ proportional to $- \mu_0 R^3 \bGzero \bcdot \bHzero$, and
 a steady force on a thin rod of radius $R$ and length $L$
 proportional to $- \mu_0 R^2 L (\bGzero \bcdot \bHzero - \tfrac{1}{2}
 (\bGzero \bcdot \bo)(\bHzero \bcdot \bo))$, where $\mu_0$ is the magnetic permeability. There
 is  torque proportional to $\mu_0 R^2 L (\bo \btimes \bHzero) (\bo \bcdot \bHzero)$ on a
 thin rod which tends to align the rod direction of the magnetic field.
 The coefficients in the force and torque expressions are functions of the
 dimensionless ratio of the radius and the penetration depth
of the magnetic field, $\beta R = \sqrt{\mu_0 \omega \kappa R^2}$,
where $\kappa$ is the electrical conductivity.
It is shown that the effect of particle
interactions can be expressed as an anisotropic diffusion term in the equation for the
particle number density. The diffusion coefficient is negative, and concentration
fluctuations are amplified, in the plane perpendicular to the magnetic field.
\end{abstract}
\maketitle
\section{Introduction}
When an electrically conducting non-magnetic particle is subject to an oscillating
magnetic field of frequency $\omega$, oscillating eddy currents are induced in the particle in accordance
with Faraday's law. An oscillating magnetic moment is induced by these eddy currents as a
result of Ampere's circuital law. The Lorentz force is the cross product of the eddy current and
the magnetic field. The interaction between the induced moment and the applied field
could result in a steady torque on a particle in a spatially uniform magnetic
field \citep{moffatt,moffatt2,vk19,vk20a}, and could also cause internal circulation
within a suspended electrically conducting drop\citep{vk24,vk25}. Although the
magnetic moment, magnetic field and eddy current are oscillating quantities with
zero average, the Lorentz force and the Maxwell stress, which are products of the current
density and the magnetic field, contain zero frequency contributions as well as
contributions with frequency $2 \omega$. The net force on an electrically conducting
particle is zero in a spatially uniform magnetic field. However, in a spatially
non-uniform oscillating field, there could be a steady force acting on symmetric particles
such as spheres and thin rods. This is due to the coupling between the instantaneous
asymmetries in the magnetic field and the eddy current distribution. The magnetophoretic
force and torque are calculated for a spherical particle and a thin rod in a general
spatially non-uniform magnetic field.

Manipulation of the trajectories of magnetic particles by magnetic fields is used
in applications such as separations \citep{bu,nameni,civelekoglu}, sorting
\citep{reiter} and drug delivery \citep{murthy}. The fundamental
principle \citep{Shao02012026,doi:10.1021/acs.langmuir.0c00839,munaz,alnaimat}
is the motion of magnetic particles due to gradients in the magnetic field.
A torque is exerted on a magnetic particle in a constant magnetic field, which tends
to align the particle with the field, but there is no net force. When a magnetic
particle is suspended in a non-magnetic medium and there is a magnetic field gradient,
there is a force acting in the
direction of increasing magnetic field based on the principle of minimisation of
the magnetic energy. This is called positive magnetophoresis. In a viscous fluid,
the particle velocity is the ratio of the magnetophoretic force and the Stokes
drag coefficient. If a non-magnetic particle is suspended in a magnetic fluid,
the particle moves in the direction of decreasing magnetic field; this is called
negative magnetophoresis. Separation is achieved either by altering the
trajectory of magnetic particles relative to non-magnetic particles, or by
capture by magnets at the walls of a conduit. Steady magnetic fields are used
in passive applications to separate magnetic particles, while active separation
involves time-dependent or rotating fields to effect separations.

There have been relatively few studies on the effect of magnetic fields on
electrically conducting particles \citep{moffatt,vk19,vk25}. There is no force
or torque on an electrically conducting non-magnetic particle in a steady
magnetic field. In a time-varying magnetic field, eddy currents are
induced in the particle due to Faraday's law of induction. These eddy currents
impart a magnetic moment to the particle due to Ampere's law, and this results
in an oscillating magnetic moment. The interaction between the oscillating magnetic
moment and the magnetic field results in a torque on a rotating particle in a uniform
magnetic field \citep{vk22}, resulting in an antisymmetric force dipole
\citep{vk19}. In the absence of relative rotation between the
particle and the field, there is a symmetric force dipole when a spherical particle
is subject to an oscillating magnetic field. For an electrically conducting drop,
the Maxwell stress \citep{vk20a} generates flow inside and outside the drop
\citep{vk25}.

Particles with a force dipole form an important part of `active matter', where
particles consume energy and self-generate motion \citep{tonertu,simharamaswamy,sriram}.
The force dipole is fixed in the particle reference frame, and the dipole translates
and rotates with the particle. The orientation vector or director of each particle
is a relevant variable in addition to the concentration and velocity fields,
and there is a particle stress in the momentum conservation equation due
to the orientation vector. The mass and momentum fluxes are formulated based
on symmetry relations, that is, the terms with lowest order in the gradients
of the field variables are included. Since this is a non-equilibrium system,
the constitutive relations contain terms of lower order in gradients compared
to those for equilibrium systems. Due to this, these systems exhibit unusual
phenomena such as long-range order, fluctuations
in number density larger than those predicted by the central limit theorem,
dynamical phase transitions and super-diffusive behaviour.

The system studied here differs from active particles in two respects.
First, the magnetic dipole moment of a spherical particle is aligned
along the magnetic field direction in a fixed reference frame and
does not rotate with the particle. For a non-spherical particle, the
magnetic moment is defined by the orientation of the particle relative
to the magnetic field direction. There has been some recent work on
collective dynamics of particles in a medium with frozen anisotropy
along one direction \citep{sriramanisotropic}. The anisotropy results in
modification of the lowest gradient terms in the formulations
for the mass and momentum fluxes, which gives rise to anisotropic
diffusion and superdiffusion.

The second difference is that, in addition
to hydrodynamic interactions, there are magnetic interactions between
particles. The constitutive relation for the flux due to interactions is
formulated by calculating the interaction force and the resulting
drift velocity in a dilute suspension. Additional terms in the constitutive
relations permitted by symmetry are not considered here, and the effective
diffusion coefficients are calculated by averaging over interactions.

A related phenomenon is electro-magneto-phoresis, where an insulating particle
in a conducting medium is subject to simultaneous electric and magnetic
fields \citep{kolin,moffatt_sellier,yariv1}. The electric and magnetic
fields are considered to be steady and the magnetic permeability
of the inclusion is the same as that of the medium. However, there is a difference
in the electrical conductivity. Due to this, there is a disturbance to the
eddy currents which generates a Lorentz force on the particle. This
phenomenon has been studied using analysis of tensorial symmetries,
boundary integral formulation and slender body theory.

Here, we consider the effect of an oscillating magnetic field on an
electrically conducting particle. The eddy currents induced in the particle
result in an oscillating magnetic moment with the same frequency as that
of the field. The Lorentz force density, which is the cross product of the current
density and the magnetic field, consists of a steady component and a component
with frequency twice that of the magnetic field. The steady component
results in a net force on a particle in a spatially non-uniform magnetic
field, and a torque on a non-spherical particle.
The phenomenon bears a resemblance to non-linear phenomena such as induced charge
electrokinetic flows \citep{levich,squiresbazant}, where an oscillating charge
distribution is induced around particles due to an oscillating electric
field, and the action of charges on the field results in steady flow.
The force and torque here are bulk phenomena, in contrast to the surface charges
induced in electroosmotic and electrophoretic flows, and are caused
by magnetic field acting on a non-magnetic but electrically conducting medium.

The force on a spherical electrically conducting particle in a spatially non-uniform
magnetic field has been calculated by Moffatt \citep{moffatt} using the Gilbert model
\citep{griffiths} for
the magnetic dipole. Here, the magnetic dipole $\bm{m}$ is considered as the superposition of two
monopoles of opposite sign and infinitesimal separation, and the force is
$\bm{F} = \mu_0 \bm{m} \bcdot \bnabla \bm{H}$, where $\mu_0$ is the magnetic
permeability and $\bm{H}$ is the magnetic field. An alternate description is the Ampere
model \citep{griffiths}, where the magnetic moment is modelled as a current loop.
In this description, the force is written as the negative of the gradient
of the potential, $\bm{F} = \mu_0 \bnabla (\bm{m} \bcdot \bm{H})$. Although the two
descriptions give the same results in most cases, there are situations such
as the hyperfine lines in a hydrogen spectrum \citep{griffiths} and the instability
in a magnetorheological suspension \citep{vk22} where the results are different; in
both cases, the Ampere description is found to be consistent with experimental
results. Here, the eddy current and Maxwell stress are first calculated,
and the force and torque are determined from the Maxwell stress distribution.

The formulation is discussed in section \ref{sec:formulation}.
Within a conducting particle subject to an oscillating field, the magnetic field
is solenoidal, and therefore it can be expressed as the curl of a potential.
Gauss's law for magnetism, Faraday's law, and Ampere's circuital law can be
combined with Ohm's law for a conducting medium to derive a Helmholtz equation
for the magnetic potential. The solution for this can be expressed in terms of
polar and spherical harmonics multiplied by the magnetic field or its gradient
using linear superposition. In the insulating medium outside the particle, the
magnetic field is irrotational and solenoidal. The magnetic field is expressed
as the gradient of a scalar potential which satisfies the Laplace equation. The
solutions are combinations of the spherical or polar harmonics and the magnetic
field or its gradient. The constants in resulting expressions are determined
using matching conditions at the surface of the particle. The force per unit
area at the surface is the dot product of the Maxwell stress and the unit
normal, and the total force and torque are calculated by integrating the
force per unit area, or its moment, over the surface.

The equations for the steady and oscillating Maxwell stress due to an oscillating
magnetic field are formulated and the notation is explained in section
\ref{sec:formulation}. The disturbance to the magnetic field due to a spherical
electrically conducting particle is examined in section \ref{sec:sph}. This calculation
requires the magnetic moment of a conducting sphere due to an oscillating magnetic
field (\cite{moffatt,landaulifshitz}), which is briefly
summarised in appendix \ref{appendix:sphere}. The magnetic moment of a spherical
particle of radius $R$ depends on the dimensionless parameter $\beta R = \sqrt{\omega
\kappa \mu_0} R$, where $\kappa$ is the electrical conductivity and $\mu_0$
is the magnetic permeability. The parameter $\beta$ is the inverse of the
penetration depth of the magnetic field into a conducting medium
(\cite{landaulifshitz}).

The steady
and oscillatory components of the Maxwell stress due to disturbance to the
magnetic field are calculated in section \ref{sec:sph}. The force is determined by
integrating the Maxwell stress over a spherical surface of radius large
compared to the particle size but small compared to the system size. Since the
particle is suspended in an insulating medium, there is no current and the
Lorentz force density is zero. Therefore, the integral of the Maxwell stress
over the particle surface is the same as that over a surface at a distance
large compared to the particle size, but small compared to the length
scale for variation of the magnetic field. This simplification is used
to calculate the force on the particle.

The disturbance to the magnetic field due to a thin rod is examined in
section \ref{sec:rod}. In this case, the oscillating magnetic moment of
a thin conducting rod is anisotropic, since the magnetic susceptibility
along the rod is different from that perpendicular to the rod. The
susceptibilities in the two directions and the magnetic moment
(\cite{landaulifshitz}) are calculated in section \ref{appendix:rod}.
The steady and oscillatory components of the Maxwell stress are calculated
in section \ref{sec:rod}. The force and torque are determined by integrating
the Maxwell stress over a spherical surface of radius large compared to the
particle length, but small compared to the length scale for the magnetic
field variation. The translation and rotation time scales in a viscous
fluid are estimated, and it is shown that rotational relaxation is fast
compared to translational motion when the rod length is much smaller than
the length scale for variation of the magnetic field.

The effect of particle interactions on the concentration evolution in a
dilute suspension is calculated by considering the effect of magnetic
interactions in section \ref{sec:particleinteractions}. In a uniform
suspension, there is no net force on a particle due to the other particles
due to symmetry. When
there is a variation in the concentration, there is a net force on a
particle due to interactions with other particles, as well as the modification
of the magnetic field due to the magnetisation of the particles. The effect
of these interactions is calculated in section \ref{subsec:particleinteractionssphere}
for spherical particles. In the long wave limit, it is shown that the effect
of interactions reduces to an anisotropic diffusion term in the concentration
equation. The diffusion coefficient parallel to the magnetic field is positive,
indicating that disturbances are damped in this direction. In contrast, the
diffusion coefficient perpendicular to the magnetic field is negative,
resulting in amplification of concentration disturbances.

The effect of interactions in a suspension of thin rods is examined in
section \ref{subsec:particleinteractionsrod}. The particle is oriented
along the stable steady orientation along the magnetic field direction.
The term arising from interactions in the particle concentration equation can
not be reduced to an anisotropic diffusion term. However, the
concentration fluctuations along the magnetic field are shown to be damped, and
fluctuations perpendicular to the magnetic field are shown to be amplified.
The significant conclusions are summarised in section \ref{sec:conclusions}.
Estimates are provided to show how the magnetophoretic force compares
to the gravitational force and how the magnetophoretic diffusion
compares to Brownian diffusion for particles of different sizes made
of conducting materials such as copper and silver.
\section{Formulation}
\label{sec:formulation}
An inclusion made of a electrically conducting neutral material in
an insulating medium is subject to an oscillating and spatially varying
magnetic field far from the inclusion,
\begin{eqnarray}
 \bH & = & (\bHzero + \bGzero \bcdot \bx) \cos{(\omega t)} \nonumber \\
 & = & \tfrac{1}{2} (\bHzero + \bGzero \bcdot \bx) (\texte^{(\imath \omega t^\dag)}+
 \texte^{- (\imath \omega t^\dag)}), \labelp{eq:01}
\end{eqnarray}
where $\omega$ is the frequency, $\imath = \sqrt{-1}$, $\bm{{\cal H}}$ is the
vector magnetic field at the center of the particle and $\bGzero$ is a second order
tensor that is the gradient of the magnetic field.
The particle has electrical conductivity $\kappa$, and the
magnetic permeability of the particle and suspending medium is considered to be $\mu_0$.
There is a disturbance to the electric and magnetic fields due to the inclusion,
\begin{eqnarray}
 \bH & = & (\tbH \texte^{(\imath \omega t)} + \tbH^{\ast} \texte^{(-\imath \omega t)}),
 \labelp{eq:02} \\
 \bE & = & (\tbE \texte^{(\imath \omega t)} + \tbE^{\ast} \texte^{(-\imath \omega t)}),
 \labelp{eq:03}
\end{eqnarray}
where $\tbH$ and $\tbE$ are complex amplitudes that depend on the spatial co-ordinates, and
the superscript $^\ast$ denotes the complex conjugates.

In the insulating medium, the Maxwell equations for the electric and magnetic field are
expressed in terms of $\tbE$ and $\tbH$,
\begin{eqnarray}
\bnabla \bcdot \tbE & = & 0, \labelp{eq:04} \\
\bnabla \btimes \tbE & = & - \mu_0 \imath \omega \tbH,
\labelp{eq:05} \\
\bnabla \bcdot \tbH & = & 0, \labelp{eq:06} \\
\bnabla \btimes \tbH & = & \cancel{\epsilon_0 \imath \omega \tbE},
\labelp{eq:07}
\end{eqnarray}
where $\epsilon_0$ is the electrical permittivity.
The last term on the right in Ampere's law, equation \ref{eq:07}, is neglected; this
approximation is valid for $(\omega R/c) \ll 1$, where $R$ is the characteristic dimension
and $c$ is the speed of light. With this approximation, the magnetic field satisfies the zero divergence
and curl conditions, $\bnabla \bcdot \tbH = \bnabla \btimes \tbH = 0$ in the
insulating medium.
Therefore, the second order tensor $\bGzero$ is symmetric and traceless,
\begin{eqnarray}
 \bGzero & = & \bGzero^T, \: \: \: \: \mbox{Trace}(\bGzero) = 0, \labelp{eq:017}
\end{eqnarray}
where $^T$ is the transpose.

In the electrically conducting inclusion, Ampere's law (equation \ref{eq:07})
is modified to incorporate the electrical current density,
\begin{eqnarray}
 \bnabla \btimes \tbH & = & \tbJ, \labelp{eq:08}
\end{eqnarray}
where the current density is given by Ohm's law,
\begin{eqnarray}
 \tbJ & = & \kappa \tbE. \labelp{eq:09}
\end{eqnarray}
If we take the curl of equation \ref{eq:08}, and use \ref{eq:09} and \ref{eq:05} for
the current density and electric field respectively, we obtain,
\begin{eqnarray}
\bnabla \btimes (\bnabla \btimes \tbH) & = & \cancel{\bnabla
(\bnabla \bcdot \tbH)} - \bnabla^{2} \tbH
= \kappa \bnabla \btimes \tbE = \imath \omega \mu_0 \kappa \tbH.
\labelp{eq:010}
\end{eqnarray}
Therefore, the equation for the magnetic field is,
\begin{eqnarray}
 \bnabla^{2} \tbH - \imath \omega \mu_0 \kappa \tbH & = & 0,
 \labelp{eq:011}
\end{eqnarray}
The magnetic field amplitude is expressed as the curl of a magnetic potential $\tbA$,
so that equation \ref{eq:06} is identically satisfied
\begin{eqnarray}
 \tbH & = & \bnabla \btimes \tbA. \labelp{eq:012}
\end{eqnarray}

The Maxwell stress tensor in the external medium is
\begin{eqnarray}
\bsigma & = & \mu_0 (\bH \bH - \tfrac{1}{2} \bI (\bH \bcdot \bH)) +
\epsilon_0 (\bE \bE - \tfrac{1}{2} \bI (\bE \bcdot \bE)),
\labelp{eq:013}
\end{eqnarray}
where $\bI$ is the identity tensor. When equations \ref{eq:02}-\ref{eq:03} are substituted
into the above expression, there are two components of the Maxwell stress, the first
with zero frequency and the second with frequency $2 \omega$,
\begin{eqnarray}
 \bsigma & = & \barbsigma + \tfrac{1}{2} (\tbsigma \exp{(2 \imath \omega t)} +
 \tbsigma^{\ast} \exp{(- 2 \imath \omega t)}), \labelp{eq:013a}
\end{eqnarray}
where
\begin{eqnarray}
 \barbsigma & = & \tfrac{1}{4} \mu_0 (\tbH \tbH^{\ast} + \tbH^{\ast} \tbH
 - \bI (\tbH \bcdot \tbH^{\ast})) \nonumber \\ & & \mbox{} + \tfrac{1}{4}
\epsilon_0 (\tbE \tbE^{\ast} + \tbE^{\ast} \tbE -
\bI (\tbE \bcdot \tbE^{\ast})), \labelp{eq:014}
\end{eqnarray}
and
\begin{eqnarray}
 \tbsigma & = & \tfrac{1}{2} \mu_0 (\tbH \tbH
 - \tfrac{1}{2} \bI (\tbH \bcdot \tbH)) \nonumber \\ & & \mbox{} + \tfrac{1}{2}
\epsilon_0 (\tbE \tbE - \tfrac{1}{2}
\bI (\tbE \bcdot \tbE)). \labelp{eq:015}
\end{eqnarray}


The equation \ref{eq:011} for the magnetic field is
\begin{eqnarray}
 \bnabla^2 \tbH + \tbe^2 \tbH & = & 0, \labelp{eq:018}
\end{eqnarray}
where $\tbe = \sqrt{- \imath} \beta$, and $\beta = \sqrt{\omega \mu_0 \kappa}$
is the inverse of the penetration depth of the magnetic
field into the conductor (\cite{landaulifshitz}).

The steady component of the Maxwell stress is
\begin{eqnarray}
  \barbsigma & = & \tfrac{1}{4} \mu_0 (\tbH \tbH^{\ast} + \tbH^{\ast} \tbH
 - \bI (\tbH \bcdot \tbH^{\ast})) \nonumber \\ & & \mbox{} + \tfrac{1}{4}
\epsilon_0 (\tbE \tbE^{\ast} + \tbE^{\ast} \tbE^{\dag} -
\bI (\tbE \bcdot \tbE^{\ast})), \labelp{eq:019}
\end{eqnarray}
and the amplitude of the oscillatory component is
\begin{eqnarray}
  \tbsigma & = & \tfrac{1}{2} \mu_0 (\tbH \tbH
 - \tfrac{1}{2} \bI (\tbH \bcdot \tbH)) \nonumber \\ & & \mbox{} + \tfrac{1}{2}
 \epsilon_0 (\tbE \tbE  - \tfrac{1}{2} \bI (\tbE \bcdot \tbE)). \labelp{eq:020}
\end{eqnarray}
The ratio of the electrical and magnetic contributions to the Maxwell stress
is estimated as follows. From Ampere's law (equation \ref{eq:09}), the current
density $|\tbJ| \sim |\tbH|/R$, and from Ohm's law (equation \ref{eq:09}),
$|\tbE| \sim |\tbJ|/\kappa \sim (|\tbH|/R \kappa)$. Therefore, the ratio of
the electrical and magnetic components of the Maxwell stress scales is
$\mu_0 |\tbH|^2/\epsilon_0 |\tbE|^2 \sim (\epsilon_0/\mu_0 \kappa^2 R^2)$.
Using typical values of electrical conductivity
$\kappa \sim 10^7 \mbox{kg}^{-1} \mbox{m}^{-3} \mbox{s}^3 \mbox{A}^2$ for
metals, magnetic permeability $\mu_0 = 4 \pi \times 10^{-7} \mbox{kg} \,
\mbox{m} \, \mbox{s}^{-2} \mbox{A}^{-2}$, and electrical permittivity
$\epsilon_0 = 8.85 \times 10^{-12} \mbox{kg}^{-1} \mbox{m}^{-3} \mbox{s}^4
\mbox{A}^2$, the ratio $(\epsilon_0/\mu_0 \kappa^2 R^2)$ is small for length
$R$ large compared to the atomic diameter. Therefore, the Maxwell stress contribution
due to the electric field is neglected in the present analysis.

The amplitude of the oscillatory component of the Maxwell
stress, \ref{eq:020}, is obtained by substituting $\tbH$ for $\tbH^\ast$ in the
expression for the steady
component of the Maxwell stress, \ref{eq:019}. Therefore, the amplitude of the
oscillatory force and torque are also obtained by the substitution $\tbH^\ast
\rightarrow \tbH$ in the resulting expression.

In the following analysis, the accent $\tilde{}$ is used to denote complex variables, while real
variables are written without the accent. The calligraphic font is used for the applied
magnetic field ($\bHzero$) and the magnetic field gradient ($\bGzero$). Bold fonts are used for
vectors and tensors, and normal fonts with subscripts are used when vectors and tensors
are expressed using indicial notation.
\section{Spherical particle}
\label{sec:sph}
 Since the curl of the magnetic field is zero outside the particle, the magnetic field
 field is expressed as the gradient of a potential, $\tbH = \bnabla \phi_H$. The potential
 satisfies the Laplace equation, $\bnabla^2 \tphi_H = 0$, because the magnetic field has
 zero divergence. The potential is a linear function of $\bHzero$ or $\bGzero$, and it
 is also a linear function of one of the spherical harmonics. The general form of the
 potential is
 \begin{eqnarray}
  \tphi_H & = & \textcolor{blue}{\bHzero \bcdot \bx} +
  \textcolor{blue}{\tfrac{1}{2} \bGzero \bddot \bx \bx} +
  \textcolor{red}{\tfrac{1}{4 \pi} \Rcu \tchi \bHzero \bcdot
  \bPhionex} + \textcolor{red}{\tfrac{1}{4 \pi} R^5 \tlambda \bGzero \bddot \bPhitwox},
  \labelp{eq:12}
 \end{eqnarray}
 and the magnetic field is,
 \begin{eqnarray}
  \tbH & = & \bnabla \tphi_H = \textcolor{blue}{\bHzero} +
  \textcolor{blue}{\bGzero \bcdot \bx} +
  \textcolor{red}{\tfrac{1}{4 \pi} \Rcu \tchi \bPhitwox \bcdot \bHzero} +
  \textcolor{red}{\tfrac{1}{4 \pi} R^5 \tlambda \bPhithreex \bddot \bGzero}, \labelp{eq:13}
 \end{eqnarray}
 where $\Rcu \tchi$ is the magnetic susceptibility, $\Rcu \tchi \bHzero$ is
 the amplitude of the induced oscillating dipole moment, $R^5 \tlambda$  is
 the susceptibility for the induced quadrupole moment and $R^5 \tlambda \bGzero$ is
 the induced quadrupole moment. The susceptibilities are defined such that $\tchi$
 and $\tlambda$ are dimensionless. These are complex constants to be evaluated
 using the boundary conditions for the magnetic field at the surface of the
 particle. In equations \ref{eq:12}, \ref{eq:13} and the following analysis,
 the blue terms are the imposed field, and the red terms are the disturbances
 due to the presence of the particle.
 The decaying harmonics $\bPhi^{(n)}$ are $n^{th}$
 order tensors which are solutions of the Laplace equation,
\begin{eqnarray}
 \bnabla^2 \bPhi^{(n)} & = & 0. \labelp{eq:14}
\end{eqnarray}
The scalar, vector and second order tensor solutions are
\begin{eqnarray}
\Phizerox & = & \frac{1}{r}, \: \: \: \bPhionex = \mbox{} - \frac{\bx}{r^3},
\: \: \: \bPhitwox = \frac{3 \bx \bx}{r^5} - \frac{\bI}{r^3}, \labelp{eq:15}
\end{eqnarray}
where $r=|\bx|$ is the distance from the particle center.
The $n^{th}$ order tensor solution, obtained by taking the gradient of
the fundamental solution $n$ times, decreases proportional to $r^{-(n+1)}$.
The spherical harmonics \ref{eq:15} are substituted into equation
\ref{eq:13}, and some simplifications are made using the properties \ref{eq:017} of
$\bGzero$, to obtain,
\begin{eqnarray}
 \tbH & = & \bHzero + \frac{\Rcu \tchi}{4 \pi}
 \left( \mbox{} - \frac{\bHzero}{r^3} + \frac{3 \bx
 (\bHzero \bcdot \bx)}{r^5} \right) + \bGzero \bcdot \bx \nonumber \\ & & \mbox{}
 + \frac{R^5 \tlambda}{4 \pi} \left( \frac{6 \bGzero \bcdot \bx}{r^5} - \frac{15 \bx
 (\bGzero \bcdot \bx)^2}{r^7} \right). \labelp{eq:15a}
\end{eqnarray}

\begin{figure}
 \includegraphics[width=.7\textwidth]{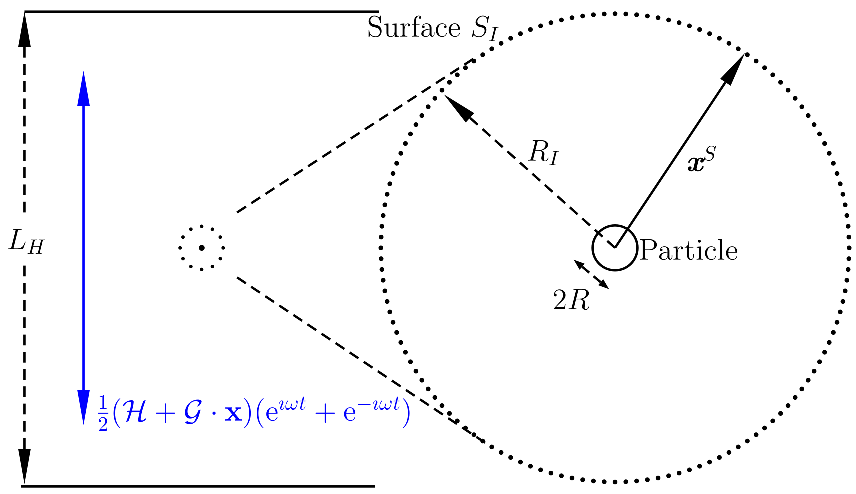}
 \caption{\label{fig:lengthsph} Schematic of the oscillating magnetic field $\bHzero$,
 the different length scales, the length
 scale $L_H$ for the magnetic field variation corresponding to the system size, the
 radius $R$ for a spherical particle and the radius $R_I$ for the spherical surface
 $S_I$, and the displacement vector $\bxs$ on this surface.}
\end{figure}

The force on the particle is usually obtained by integrating the Maxwell stress
over the particle surface. However, it is possible to integrate over any surface
in the medium surrounding the particle, since the current density and the
Maxwell force density in the surrounding medium are zero. Therefore, the net force
and torque calculated over any surface that encloses the particle
is equal to that exerted on the particle. There are two length
scales in the problem, the particle radius $R$ and the length scale $L_H$ for
the variation of the magnetic field. In the Taylor expansion \ref{eq:01} for the
magnetic field, it is implicitly assumed that $R \ll L_H$. The force
and torque are calculated by integrating over an intermediate surface $S_I$ of
radius $R_I$, where the length scale $R_I$ is much larger than the particle size
but much smaller than the length scale $L_H$ of the magnetic field, as shown in figure
\ref{fig:lengthsph}.

When the distance from the particle is comparable to $R_I$, the magnitude of
the magnetic field gradient times the distance from the particle is small
compared to the magnitude of the magnetic field, $|\bxs \bcdot \bGzero| \ll |\bHzero|$.
The disturbance to the magnetic field due to the particle scales as a
power of $(R/r)$, the ratio of the radius and distance from the center.
The integral over the surface at $S_I$ is finite in the limit $R_I \gg R$
only for an integrand proportional to $(R/r)^2$, since the product of the
integrand and the surface area is finite.

The steady and oscillatory forces are calculated by integrating the Maxwell
stress over this surface,
\begin{eqnarray}
 \barbF & = & \tfrac{1}{4} \mu_0 \int_{S_I} \total S_I \left( \tbH \tbH^\ast +
 \tbH^\ast \tbH - \bI \tbH \bcdot \tbH^\ast \right) \bcdot \bns, \labelp{eq:16}
\end{eqnarray}
where $S_I$ is the spherical surface at a distance $R_I$ from the particle,
$\bns = (\bxs/R_I)$ is the outward unit normal, and $\bxs$ is a location on
the surface $S_I$, as shown in figure \ref{fig:lengthsph}.

The terms in equation \ref{eq:16} are simplified as follows. The integral
of the first term in the brackets on the right dotted with the unit normal is
\begin{eqnarray}
 \int_{S_I} \total S_I \tbH \tbH^\ast \bcdot \bns & = &
 \int_{S_I} \total S_I \left( \textcolor{blue}{\bHzero} +
  \textcolor{blue}{\bGzero \bcdot \bxs} +
  \textcolor{red}{\tfrac{1}{4 \pi} \Rcu \tchi \bPhitwos \bcdot \bHzero} \right.
  \nonumber \\ & & \mbox{} \left. +
  \textcolor{red}{\tfrac{1}{4 \pi} R^5 \tlambda \bPhithrees \bddot \bGzero} \right)
  \left( \textcolor{blue}{\bHzero} +
  \textcolor{blue}{\bGzero \bcdot \bxs} \right.  \nonumber \\
  & & \left. \mbox{} \: \: +
  \textcolor{red}{\tfrac{1}{4 \pi} \Rcu \tchia \bPhitwos \bcdot \bHzero} +
  \textcolor{red}{\tfrac{1}{4 \pi} R^5 \tlambdaa \bPhithrees \bddot \bGzero}
  \right) \bcdot \bns. \labelp{eq:17}
\end{eqnarray}
The product of two blue terms in the integral \ref{eq:17} is the stress due to the
imposed field in the absence of the particle. It can be shown that the integrals
of these terms are zero. The product of two red terms in
the integral \ref{eq:17}, decreases as a higher power of $(1/r)$ than the product
of one red and one blue term in the limit $R_I \gg R$. Therefore, the largest
contribution is due to the product of a blue and a red terms,
\begin{eqnarray}
 \int_{S_I} \total S_I \tbH \tbH^\ast \bcdot \bns & = &
 \int_{S_I} \total S_I \left( \cancel{\textcolor{blue}{\bHzero}
 (\textcolor{red}{\tfrac{1}{4 \pi} \Rcu \tchia \bPhitwos \bcdot \bHzero})} +
 \overset{{\tiny \textcircled{1}}}{\textcolor{blue}{\bHzero} (\textcolor{red}{
 \tfrac{1}{4 \pi} R^4 \tlambdaa\bPhithrees \bddot \bGzero})} \right. \nonumber \\
 & & \left. \mbox{}
 + \overset{{\tiny \textcircled{2}}}{
 (\textcolor{blue}{\bGzero \bcdot \bxs}) (\textcolor{red}{\tfrac{1}{4 \pi} \Rcu \tchia
 \bPhitwos \bcdot \bHzero})} \right. \nonumber \\ & & \left. \mbox{}
 + \cancel{(\textcolor{blue}{\bGzero \bcdot \bxs}) (\textcolor{red}{\tfrac{1}{4 \pi}
 R^4 \tlambdaa \bPhithrees \bddot \bGzero})}
 + \cancel{(\textcolor{red}{\tfrac{1}{4 \pi} \Rcu \tchi \bPhitwos \bcdot \bHzero})\textcolor{blue}{\bHzero}}
 \right. \nonumber \\ & & \left. \mbox{}
 + \overset{{\tiny \textcircled{3}}}{(\textcolor{red}{\tfrac{1}{4 \pi} R^4 \tlambda
 \bPhithrees \bddot \bGzero}) \textcolor{blue}{\bHzero}}
 + \overset{{\tiny \textcircled{4}}}{(\textcolor{red}{\tfrac{1}{4 \pi}
 \Rcu \tchi \bPhitwos \bcdot \bHzero}) (\textcolor{blue}{\bGzero \bcdot \bxs})}
 \right. \nonumber \\ & & \left. \mbox{} + \cancel{(\textcolor{red}{\tfrac{1}{4 \pi} \tlambda
 \bPhithrees \bddot \bGzero}) (\textcolor{blue}{\bGzero \bcdot \bxs})}
 \right) \bcdot \bns. \labelp{eq:18}
\end{eqnarray}
In the above equation, the cancelled terms,
when multiplied by the unit normal $\bns$, are odd functions of $\bxs$; when odd functions
of $\bxs$ are integrated over a spherical surface, the result is zero. Note that
$\bns$ is an odd function of $\bxs$, and $\bPhitwos$ and $\bPhithrees$ are even and
odd functions of $\bxs$ respectively. The even functions of $\bx$ in equation \ref{eq:18}
are numbered for ease of discussion. The terms $\textcircled{1}$ and $\textcircled{3}$
decrease proportional to $r^{-4}$ for $r \sim R_I$, and the surface area increases
proportional to $r^2$. Therefore, the integrals of these terms tend to
zero for $R_I \gg R$. The terms $\textcircled{2}$ and $\textcircled{4}$
decrease proportional to $r^{-2}$, and the surface area increases proportional
to $r^2$. Therefore, the terms $\textcircled{2}$ and $\textcircled{4}$ provide
a finite contribution to the integral over the surface at $R_I$. This results in the
following simplification of equation \ref{eq:18},
\begin{eqnarray}
  \int_{S_I} \total S_I \tbH \tbH^\ast \bcdot \bns & = &
  \int_{S_I} \total S_I [\textcolor{blue}{(\bGzero \bcdot \bxs)}
  \textcolor{red}{(\tfrac{1}{4 \pi} \Rcu \tchia \bPhitwos
  \bcdot \bHzero)} \bcdot \bns \nonumber \\ & & \mbox{}
  + \textcolor{red}{(\tfrac{1}{4 \pi} \Rcu \tchi \bPhitwos \bcdot \bHzero)}
  \textcolor{blue}{(\bGzero \bcdot \bxs)} \bcdot \bns]. \labelp{eq:19}
\end{eqnarray}

The second term in the brackets in the integrand in \ref{eq:16} is
the complex conjugate of the first term. The third term in the brackets in the
integrand, when dotted with the unit normal and integrated over the surface is
\begin{eqnarray}
   \int_{S_I} \total S_I (\mbox{} \tbH \bcdot \tbH^\ast) \bns & = &
  \int_{S_I} \total S_I [\textcolor{blue}{(\bGzero \bcdot \bxs)} \bcdot
  \textcolor{red}{(\tfrac{1}{4 \pi} \Rcu \tchia \bPhitwos \bcdot \bHzero)}
  \nonumber \\ & & \mbox{}
  + \textcolor{red}{(\tfrac{1}{4 \pi} \Rcu \tchi \bPhitwos \bcdot \bHzero)} \bcdot
  \textcolor{blue}{(\bGzero \bcdot \bxs)}] \bns. \labelp{eq:20}
\end{eqnarray}
Here, the simplification procedures adopted are identical to those in going
from equation \ref{eq:17} to \ref{eq:19}.

The integrals in equations \ref{eq:19} and \ref{eq:20} are evaluated using indicial notation.
The integral in equation \ref{eq:19} is
\begin{eqnarray}
  \int_{S_I} \total S_I \tH_i \tH_j^\ast \ns_j & = &
  \textcolor{red}{\tfrac{1}{4 \pi} \Rcu \tchia \Hzero_l}
  \textcolor{blue}{\Gzero_{ik}}  \int_{S_I} \total S_I \textcolor{blue}{\xs_k}
  \textcolor{red}{\Phitwo_{lj}(\bxs)} \times (\xs_j/r) \nonumber \\ & & \mbox{}
  + \textcolor{red}{\tfrac{1}{4 \pi} \Rcu \tchi \Hzero_k}
  \textcolor{blue}{\Gzero_{jl}} \int_{S_I} \total S_I
  \textcolor{red}{\Phitwo_{ik}(\bxs)} (\textcolor{blue}{\xs_l} \xs_j/r)] \nonumber \\
  & = & \Rcu (\textcolor{red}{\tchia \Hzero_l}
  \textcolor{blue}{\Gzero_{ik}} (\tfrac{2}{3} \delta_{lk})+
  \textcolor{red}{\tchi \Hzero_k} \textcolor{blue}{\Gzero_{jl}}
  \left(\tfrac{1}{5} (\delta_{il} \delta_{jk} + \delta_{ij} \delta_{lk})
  - \tfrac{2}{15} \delta_{ik} \delta_{jl} \right)) \nonumber \\
    & = &  \Rcu (\textcolor{blue}{\Gzero_{ik}}
    \textcolor{red}{(\tfrac{2}{3} \tchia + \tfrac{1}{5} \tchi) \Hzero_k}
    + \tfrac{1}{5} \textcolor{blue}{\Gzero_{ki}} \textcolor{red}{\tchi \Hzero_k}
    - \tfrac{2}{15} \delta_{ij} \textcolor{blue}{\Gzero_{kk}} \textcolor{red}{
    \tchi \Hzero_i}).
  \labelp{eq:21}
\end{eqnarray}
Here, we have used the expression \ref{eq:15} for $\bPhitwos$, and the identities
\begin{eqnarray}
 \tfrac{1}{4 \pi} \int_{S_I} \total S_I \xs_i \xs_j & = & \tfrac{1}{3} \delta_{ij} R_I^4,
 \labelp{eq:21a} \\
 \tfrac{1}{4 \pi} \int_{S_I} \total S_I \xs_i \xs_j \xs_k \xs_l & = & \tfrac{1}{15}
 (\delta_{ij} \delta_{kl} + \delta_{ik} \delta_{jl} + \delta_{il} \delta_{jk}) R_I^6.
 \labelp{eq:21b}
\end{eqnarray}

The integral in equation \ref{eq:20} is evaluated in a similar manner,
\begin{eqnarray}
  \int_{S_I} \total S_I \tH_j \tH_j^\ast \ns_i & = &
  \textcolor{red}{\tfrac{1}{4 \pi} \Rcu \tchia \Hzero_l} \textcolor{blue}{\Gzero_{jk}}
  \int_{S_I} \total S_I \textcolor{blue}{\xs_k} \textcolor{red}{\Phitwo_{lj}(\bxs)} \times (\xs_i/r)
  \nonumber \\ & & \mbox{}
  + \textcolor{red}{\tfrac{1}{4 \pi} \Rcu \tchi \Hzero_k} \textcolor{blue}{\Gzero_{jl}} \int_{S_I} \total S_I
  \textcolor{red}{\Phitwo_{jk}(\bxs)} \times (\textcolor{blue}{\xs_l} \xs_i/r)] \nonumber \\ & = & \Rcu (
  \textcolor{red}{\tchia \Hzero_l} \textcolor{blue}{\Gzero_{jk}} (\tfrac{1}{5} (\delta_{ij} \delta_{kl} +
  \delta_{il} \delta_{jk}) - \tfrac{2}{15} \delta_{ik} \delta_{jl})
  \nonumber \\ & & \mbox{} +
  \textcolor{red}{\tchi \Hzero_k} \textcolor{blue}{\Gzero_{jl}} (\tfrac{1}{5}(\delta_{ij} \delta_{kl} +
  \delta_{ik} \delta_{jl}) - \tfrac{2}{15} \delta_{il} \delta_{jk}))
  \nonumber \\
& = & \Rcu (\tfrac{1}{5} \textcolor{blue}{\Gzero_{ik}} \textcolor{red}{\Rcu (\tchi + \tchia) \Hzero_k} +
\tfrac{1}{5} \textcolor{blue}{\Gzero_{kk}} \textcolor{red}{(\tchi + \tchia) \Hzero_i}
\nonumber \\ & & \mbox{} -
\tfrac{2}{15} \textcolor{blue}{\Gzero_{ki}} \textcolor{red}{(\tchi + \tchia) \Hzero_k}).
\labelp{eq:22}
\end{eqnarray}
The symmetric and traceless nature of $\bGzero$,  $\Gzero_{jk} = \Gzero_{kj}$ and
$\Gzero_{kk} = 0$ are used to simplify the equations \ref{eq:21} and \ref{eq:22},
\begin{eqnarray}
   \int_{S_I} \total S_I \tH_i \tH_j^\ast \ns_j & = &
   \Rcu \textcolor{blue}{\Gzero_{ik}} \textcolor{red}{(\tfrac{2}{3} \tchia + \tfrac{2}{5} \tchi)
   \Hzero_k}, \label{eq:21a}\\
  \int_{S_I} \total S_I \tH_j \tH_j^\ast \ns_i
    & = & \tfrac{1}{15} \Rcu \textcolor{blue}{\Gzero_{ik}} \textcolor{red}{(\tchi + \tchia) \Hzero_k}. \labelp{eq:22a}
\end{eqnarray}
The results \ref{eq:21a} and \ref{eq:22a} are substituted in the expression for the steady force \ref{eq:16},
\begin{eqnarray}
 \barF_i & = &
 \tfrac{1}{4} \mu_0 \textcolor{blue}{\Gzero_{ij}} \textcolor{red}{\Rcu (\tchi + \tchia) \Hzero_j}. \labelp{eq:23}
\end{eqnarray}

The steady second order force moment is,
\begin{eqnarray}
\barbK & = & \tfrac{1}{4} \mu_0 \int_{S_I} \total S_I \bxs (\tbH \tbH^\ast + \tbH^\ast \tbH - \bI
\tbH \bcdot \tbH^\ast) \bcdot \bns. \labelp{eq:25}
\end{eqnarray}
Using equation \ref{eq:13} for $\tbH$, the contribution due to the first term in the above
equation is
\begin{eqnarray}
 \int_{S_I} \total S_I \bxs \tbH \tbH^\ast \bcdot \bns
 & = & \tfrac{1}{4} \mu_0 \int_{S_I} \total S_I \bxs
 \left( \textcolor{blue}{\bHzero} +
  \textcolor{blue}{\bGzero \bcdot \bxs} +
  \textcolor{red}{\tfrac{1}{4 \pi} \Rcu \tchi \bPhitwos \bcdot \bHzero} \right. \nonumber \\ & &
  \mbox{} \left. +
  \textcolor{red}{\tfrac{1}{4 \pi} \tlambda
  \bPhithrees \bddot \bGzero} \right)
  \left( \textcolor{blue}{\bHzero} +
  \textcolor{blue}{\bGzero \bcdot \bxs} \right. \nonumber \\ & & \left. \mbox{} +
  \textcolor{red}{\tfrac{1}{4 \pi} \Rcu \tchia \bPhitwos \bcdot \bHzero} +
  \textcolor{red}{\tfrac{1}{4 \pi} \tlambdaa
  \bPhithrees \bddot \bGzero} \right)\bcdot \bns. \labelp{eq:26}
\end{eqnarray}
The quadratic product of the blue terms in the above equation is due to the
oscillating field in the absence of the particle. The largest
contribution to the particle force moment is due to the product of the red and blue terms,
\begin{eqnarray}
 \int_{S_I} \total S_I \bxs \tbH \tbH^\ast \bcdot \bns & = &
 \int_{S_I} \total S_I \bxs \left( \overset{{\tiny \textcircled{1}}}{\textcolor{blue}{\bHzero}
 (\textcolor{red}{\tfrac{1}{4 \pi} \Rcu \tchia \bPhitwos \bcdot \bHzero})} +
 \cancel{\textcolor{blue}{\bHzero} (\textcolor{red}{\tfrac{1}{4} \tlambdaa
 \bPhithrees \bddot \bGzero})} \right. \nonumber \\ & & \left. \mbox{} +
 \cancel{(\textcolor{blue}{\bGzero \bcdot \bxs}) (\textcolor{red}{\tfrac{1}{4 \pi} \Rcu \tchia
 \bPhitwos \bcdot \bHzero})} \right. \nonumber \\ & & \left. \mbox{} +
 \overset{{\tiny \textcircled{2}}}{(\textcolor{blue}{\bGzero \bcdot \bxs}) (\textcolor{red}{
 \tfrac{1}{4 \pi} \tlambdaa \bPhithrees \bddot \bGzero})}
 + \overset{{\tiny \textcircled{3}}}{(\textcolor{red}{\tfrac{1}{4 \pi} \Rcu \tchi \bPhitwos
 \bcdot \bHzero})\textcolor{blue}{\bHzero}} \right. \nonumber \\ & & \left. \mbox{}
 + \cancel{(\textcolor{red}{\tfrac{1}{4 \pi} \tlambda
 \bPhithrees \bddot \bGzero}) \textcolor{blue}{\bHzero}}
 + \cancel{(\textcolor{red}{\tfrac{1}{4 \pi} \Rcu \tchi \bPhitwos \bcdot \bHzero})
 (\textcolor{blue}{\bGzero \bcdot \bxs})}  \right. \nonumber \\ & & \left. \mbox{}
 + \overset{{\tiny \textcircled{4}}}{(\textcolor{red}{\tfrac{1}{4 \pi}
 \tlambda \bPhithrees \bddot \bGzero}) (\textcolor{blue}{\bGzero \bcdot \bxs})}
 \right) \bcdot \bns. \labelp{eq:27}
\end{eqnarray}
Here, the cancelled terms (multiplied by $\bxs$ and dotted with $\bns$) are odd functions of $\bxs$,
and therefore the integrals are zero. The product of $\bxs$ and the terms $\textcircled{1}$-$\textcircled{4}$
decrease proportional to $r^{-2}$ for $r \sim R_I$, and the surface area increases proportional to
$r^2$. Therefore, the integrals of these terms over the surface $S_I$ are finite.
However, the terms $\textcircled{2}$ and $\textcircled{4}$ are higher order in gradients compared
to $\textcircled{1}$ and $\textcircled{3}$, and therefore are neglected. The integral of the
first term in the brackets in equation \ref{eq:27} multiplied by $\bxs$ and dotted with $\bns$
is evaluated using indicial notation and the identities \ref{eq:21a} and \ref{eq:21b},
\begin{eqnarray}
 \int_{S_I} \total S_I \xs_i \tH_j \tH_k^\ast \ns_k & = &
 \int_{S_I} \total S_I \left( \xs_i \textcolor{blue}{\Hzero_j} \textcolor{red}{(\tfrac{1}{4 \pi}
 \Rcu \tchia \Phitwo_{kl}(\bxs) \Hzero_l)} (\xs_k/r) \right. \nonumber \\ & & \left. \mbox{} +
  x_i  \textcolor{red}{(\tfrac{1}{4 \pi} \Rcu \tchi \Phitwo_{jl}(\bxs) \Hzero_l)}
  \textcolor{blue}{\Hzero_k} (\xs_k/r) \right) \nonumber \\
  & = & \Rcu (\textcolor{blue}{\Hzero_j} \textcolor{red}{\tchia \Hzero_l} (\tfrac{2}{3} \delta_{il}) +
  \textcolor{red}{\tchi \Hzero_l} \textcolor{blue}{\Hzero_k} (\tfrac{1}{5} (\delta_{ij} \delta_{kl} +
  \delta_{il} \delta_{jk}) - \tfrac{2}{15} \delta_{ik} \delta_{jl})) \nonumber \\ 
 & = &
    \Rcu (\tfrac{2}{3} \textcolor{red}{\tchia \Hzero_i} \textcolor{blue}{\Hzero_j} +
    \tfrac{1}{5} \textcolor{red}{\tchi \Hzero_i} \textcolor{blue}{\Hzero_j} +
    \tfrac{1}{5} \delta_{ij} \textcolor{red}{\tchi \Hzero_k} \textcolor{blue}{\Hzero_k}
     \nonumber \\ & & \mbox{}
    - \tfrac{2}{15} \textcolor{blue}{\Hzero_i} \textcolor{red}{\tchi \Hzero_j}). \labelp{eq:29}
\end{eqnarray}
 The integral of the third term in the brackets in equation \ref{eq:25} multiplied
 by $\bxs$ and dotted with $\bns$ is
\begin{eqnarray}
 \int_{S_I} \total S_I \xs_i \tH_k \tH_k^\ast \ns_j & = &
 \int_{S_I} \total S_I \left( \xs_i \textcolor{blue}{\Hzero_k} \textcolor{red}{(\tfrac{1}{4 \pi} \Rcu \tchia
 \Phitwo_{kl}(\bxs) \Hzero_l)} (\xs_j/r) \right. \nonumber \\ & & \left. \mbox{} +
 \xs_i \textcolor{red}{(\tfrac{1}{4 \pi} \Rcu \tchi \Phitwo_{kl}(\bxs) \Hzero_l)} \textcolor{blue}{\Hzero_k} (\xs_j/r) \right)
 \nonumber \\
 & = & \Rcu (\textcolor{blue}{\Hzero_k} \textcolor{red}{\tchia \Hzero_l}
 + \textcolor{red}{\tchi \Hzero_l} \textcolor{blue}{\Hzero_k} (\tfrac{1}{5} (\delta_{ik} \delta_{jk}
 + \delta_{il} \delta_{jk}) - \tfrac{2}{15} \delta_{ij} \delta_{kl})) \nonumber \\
 & = & \Rcu (\tfrac{1}{5} \textcolor{blue}{\Hzero_i} \textcolor{red}{(\tchi + \tchia) \Hzero_j} +
 \tfrac{1}{5} \textcolor{red}{(\tchi + \tchia) \Hzero_i} \textcolor{blue}{\Hzero_j}
 \nonumber \\ & & \mbox{} -
 \tfrac{2}{15} \delta_{ij} \textcolor{blue}{\Hzero_k} \textcolor{red}{(\tchi + \tchia)
 \Hzero_k}). \labelp{eq:291}
 \end{eqnarray}

The expression \ref{eq:29}, its complex conjugate and \ref{eq:291} are substituted into the expression
\ref{eq:25} to obtain,
\begin{eqnarray}
 \barK_{ij} & = & \tfrac{1}{4} \mu_0 R^3 \left( \tfrac{2}{3} \textcolor{red}{(\tchi + \tchia) \Hzero_i}
 \textcolor{blue}{\Hzero_j} - \tfrac{1}{3} \textcolor{blue}{\Hzero_i}
 \textcolor{red}{(\tchi + \tchia) \Hzero_j} + \tfrac{1}{3} \delta_{ij}
 \textcolor{blue}{\Hzero_k} \textcolor{red}{(\tchi + \tchia) \Hzero_k} \right) \nonumber \\
 & = &
\tfrac{1}{12} \mu_0 \Rcu (\tchi + \tchia)(\Hzero_i \Hzero_j + \delta_{ij} \Hzero_k \Hzero_k).
 \labelp{eq:293}
\end{eqnarray}

The expressions \ref{eq:23} and \ref{eq:293} are expressed in vector notation,
\begin{eqnarray}
 \barbF & = &
  \mbox{} - \mu_0 \Rcu \barGamma_s \bGzero \bcdot \bHzero, \labelp{eq:24}
\end{eqnarray}
\begin{eqnarray}
 \barbK & = & \mbox{} - \tfrac{1}{3} \mu_0 \Rcu \barGamma_s (\bHzero \bHzero +
 \bI \bHzero \bcdot \bHzero), \labelp{eq:294}
\end{eqnarray}
where
\begin{eqnarray}
 \barGamma_s = \mbox{} - \tfrac{1}{4} (\tchi + \tchia). \labelp{eq:296}
\end{eqnarray}

In addition to the steady parts of the force and the force moment, there is an
oscillatory component with frequency $2 \omega$. Comparing the expressions
\ref{eq:019} and \ref{eq:020}, it is easily inferred that the amplitude of
the oscillatory component is obtained by the substitution $\tbH^\ast \rightarrow
\tbH$ in the steady part. This is equivalent to the substitution $\tchi \rightarrow
\tchia$ in the expressions \ref{eq:24} and \ref{eq:294},
\begin{eqnarray}
 \tbF & = & 
 \mbox{} - \mu_0 \Rcu \tGamma_s \bGzero \bcdot \bHzero. \labelp{eq:295}
\end{eqnarray}
\begin{eqnarray}
 \tbK & = & 
 \mbox{} - \tfrac{1}{3} \mu_0 \Rcu \tGamma_s (\bHzero \bHzero +
  \tfrac{1}{2} \bI \bHzero \bcdot \bHzero), \labelp{eq:297a}
\end{eqnarray}
where
\begin{eqnarray}
 \tGamma_s = - \tfrac{1}{2} \tchi. \labelp{eq:297}
\end{eqnarray}
Comparing equations \ref{eq:296} and \ref{eq:297}, it is evident that
$\barGamma_s = \mbox{Re}(\tGamma_s)$.

\begin{figure}
\psfrag{x}[][][1][0]{{$\beta R$}}
\psfrag{y}[][][1][0]{\textcolor{blue}{$\barGamma_s
=\mbox{Re}(\tGamma_s)$}, \textcolor{red}{$\mbox{Im}(\tGamma_s)$}}

 \includegraphics[width=.49\textwidth]{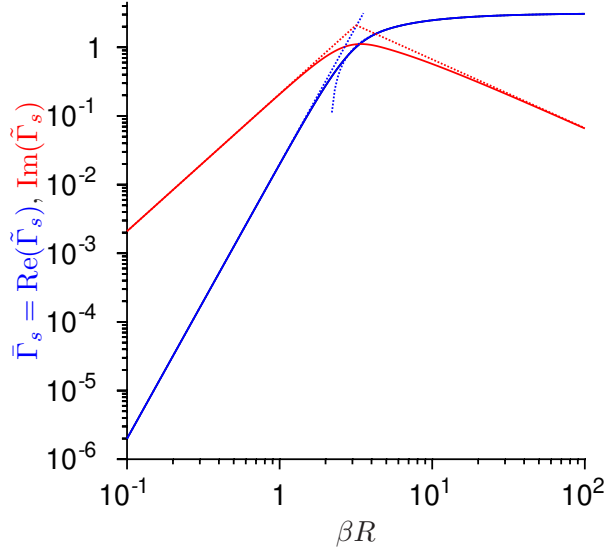}

 \caption{\label{fig:fieldgradient1} The coefficients $\barGamma_s = \mbox{Re}(\tGamma_s)$ and
 $\mbox{Im}(\tGamma_s)$ as a function of $\beta R$.
 The dashed blue lines on the left and right are
 the asymptotic results \ref{eq:225} and \ref{eq:226} respectively, and the dashed red lines on
 the left and right are the asymptotic results \ref{eq:229} and \ref{eq:230} respectively.}
\end{figure}

The susceptibility $\tchi$ is calculated in appendix \ref{appendix:sphere}
using the procedures of \cite{moffatt} and \cite{landaulifshitz},
\begin{eqnarray}
\tchi & = & \mbox{} - 2 \pi \left(\mbox{} 1 - \frac{3}{(\tbe R)^2} +
\frac{3 \cot{(\tbe R)}}{\tbe R} \right).
\labelp{eq:298}
\end{eqnarray}
This is substituted in equations \ref{eq:296} and \ref{eq:297} to obtain the variation of
$\barGamma_s$ and $\tGamma_s$ with the dimensionless parameter $\beta R$. These
coefficients are shown in figure \ref{fig:fieldgradient1}.

The asymptotic
behaviour of $\barGamma_s$ and $\tGamma_s$ for $\beta R \ll 1$ and $\beta R \gg 1$ are
\begin{eqnarray}
 \barGamma_s = \mbox{Re}(\tGamma_s) & = & \mbox{} \frac{2 \pi (\beta R)^4}{315} \: \: \: \mbox{for} \: \: \:
 \beta R \ll 1, \labelp{eq:225} \\ & = &
 \mbox{} \pi \left( 1 - \frac{3}{\sqrt{2} \beta R} \right)
 \: \: \: \mbox{for} \: \: \:
 \beta R \gg 1, \labelp{eq:226}
\end{eqnarray}
\begin{eqnarray}
 \mbox{Im}(\tGamma_s) & = & \mbox{} \frac{(\beta R)^2}{15} \: \: \: \mbox{for} \: \: \:
 \beta R \ll 1, \labelp{eq:229} \\ & = &
 \mbox{} \frac{3 \pi}{\sqrt{2} \beta R}
 \: \: \: \mbox{for} \: \: \: \beta R \gg 1. \labelp{eq:230}
\end{eqnarray}
The coefficient $\barGamma_s$ increases proportional to $(\beta R)^4$ for
$(\beta R) \ll 1$, and tends to a constant value for $\beta R \gg 1$.
The coefficient $\mbox{Im}(\tGamma_s)$ increases proportional to
$(\beta R)^2$ for $\beta R \ll 1$, and decreases proportional to
$(\beta R)^{-1}$ for $\beta R \gg 1$. Thus, the oscillatory response
has the same phase as the applied magnetic field for $\beta R \gg 1$,
whereas there is a phase shift by $\pi/2$ for $\beta R \ll 1$.
For $\barGamma_s$, figure \ref{fig:fieldgradient1} shows that the $\beta R \ll 1$
approximation \ref{eq:225} is valid for $\beta R \lesssim 1$, while the approximation
\ref{eq:226} is valid for $\beta R \gtrsim 2$. For $\mbox{Im}(\tGamma_s)$, the
approximation \ref{eq:229} is valid for $\beta R \sim 1$, while the approximation
\ref{eq:230} is valid for $\beta R \gtrsim 10$.


The result \ref{eq:294} provides the steady force on a spherical particle in three dimensions.
Therefore, particle migration is towards locations where the gradient of the applied
field is zero.
The simplest pattern of field lines that satisfy zero divergence and zero curl are the
quadrupolar growing harmonics in three dimensions which is expressed in Cartesian
coordinates as
\begin{eqnarray}
 \tH_x & = & \gamma_x x, \: \: \: \tH_y = \gamma_y y, \: \: \: \tH_z = \gamma_z z. \labelp{eq:42}
\end{eqnarray}
The coefficients $\gamma_x, \gamma_y$ and $\gamma_z$ are constrained by the
zero divergence condition $\gamma_x + \gamma_y + \gamma_z = 0$. In this field, the force is,
\begin{eqnarray}
\tbF & = & \mbox{} - \mu_0 R^3 \barGamma_s (\gamma_x^2 x \bme_x + \gamma_y^2 y \bme_y
+ \gamma_z^2 z \bme_z). \labelp{eq:43}
\end{eqnarray}
Thus, a spherical particle migrates to the origin in this general quadrupolar
field.


 \section{Thin rod}
\label{sec:rod}
\begin{figure}
 \includegraphics[width=.7\textwidth]{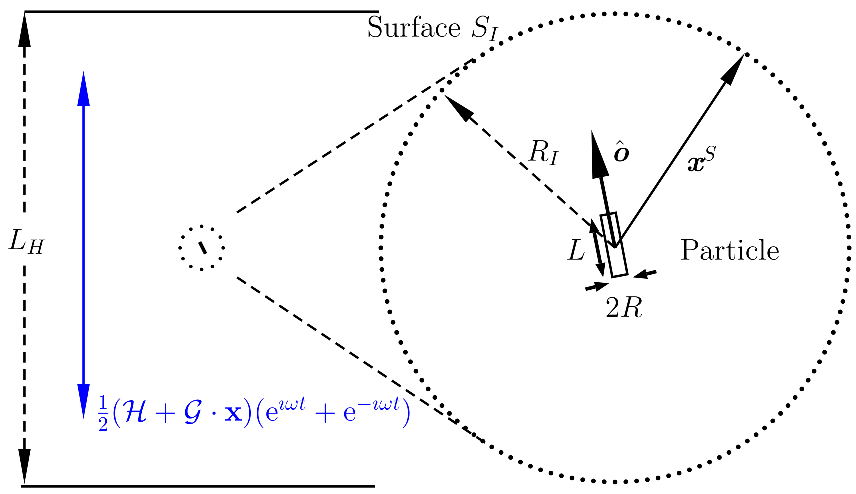}
 \caption{\label{fig:lengthrod} Schematic of the oscillating magnetic field $\bHzero$,
 the different length scales, thin rod of radius $R$ and length $L$ and orientation
 vector $\bo$, of large aspect ratio, $L \gg R$. Also shown are the length
 scale $L_H$ for the magnetic field variation corresponding to the system size, the
 radius $R_I>>L$ for the spherical surface
 $S_I$, and the displacement vector $\bxs$ on this surface.}
\end{figure}

In contrast to a sphere, the magnetic dipole moment for a thin rod depends on the relative
orientation between the axis $\bo$ and the magnetic field, as shown in figure
\ref{fig:lengthrod}. That is, the rod has different
susceptibilities in the directions parallel and perpendicular to the axis. The
magnetic field is resolved into two components, the longitudinal component
$\bo (\bo \bcdot \bHzero)$ parallel to the axis and the
transverse component $(\bI - \bo \bo) \bcdot \bHzero$ perpendicular to
the axis. The equivalent of equation \ref{eq:12} for the magnetic potential is
 \begin{eqnarray}
  \tphi_H & = & \textcolor{blue}{\bHzero \bcdot \bx} +
  \textcolor{blue}{\tfrac{1}{2} \bGzero \bddot \bx \bx} +
  \textcolor{red}{\tfrac{1}{4 \pi} \RsqL \tchipar \bo (\bo \bcdot \bHzero) \bcdot \bPhionex}
  \nonumber \\ & & \hspace{.5in} \mbox{}
  + \textcolor{red}{\tfrac{1}{4 \pi} \RsqL \tchiperp ((\bI - \bo \bo) \bcdot \bHzero) \bcdot \bPhionex},
  \labelp{eq:31}
 \end{eqnarray}
 and the equivalent of equation \ref{eq:13} for the magnetic field is
 \begin{eqnarray}
  \tbH & = & \bnabla \tphi_H = \textcolor{blue}{\bHzero} +
  \textcolor{blue}{\bGzero \bcdot \bx} +
  \textcolor{red}{\tfrac{1}{4 \pi} \RsqL \tchipar \bPhitwox \bcdot \bo (\bo \bcdot \bHzero)}
  \nonumber \\ & & \hspace{.5in} \mbox{}
  + \textcolor{red}{\tfrac{1}{4 \pi} \RsqL \tchiperp \bPhitwox \bcdot ((\bI - \bo \bo) \bcdot \bHzero)},
  \labelp{eq:32}
 \end{eqnarray}
 where $\RsqL \tchipar$ and $\RsqL \tchiperp$ are the magnetic susceptibilities parallel
 and perpendicular to the cylinder axis. These susceptibilities are calculated in appendix
 \ref{appendix:rod} using the procedure of \cite{landaulifshitz}.
Here, we have neglected the equivalents of the terms proportional to $\tlambda$ in
equations \ref{eq:12} and \ref{eq:13}, since the analysis in section
\ref{sec:sph} has shown that these do not contribute to the force
and the force moment.

The analysis is carried out using the same procedure as in section
\ref{sec:sph}. The net force is calculated by integrating the Maxwell
stress over the spherical surface $S_I$  in figure \ref{fig:lengthrod}. The radius
$R_I$ of the sphere is much larger than the length of the rod, but much smaller
than the length scale $L_H$ for the variation of the magnetic field.

It is evident that equations \ref{eq:31} and
\ref{eq:32} are obtained by the transformations $\textcolor{red}{\Rcu \tchi \bHzero} \rightarrow
\textcolor{red}{\RsqL \tchipar \bo (\bo \bcdot \bHzero) + \RsqL \tchiperp (\bI - \bo \bo) \bcdot \bHzero}$
and $\textcolor{red}{\Rcu \tchia \bHzero} \rightarrow
\textcolor{red}{\RsqL \tchipara \bo (\bo \bcdot \bHzero) + \RsqL \tchiperpa (\bI - \bo \bo) \bcdot \bHzero}$
in equations \ref{eq:12} and \ref{eq:13}.
Therefore, the force is obtained using these same transformations in equation \ref{eq:23},
\begin{eqnarray}
 \barbF & = & \tfrac{1}{4} \mu_0 [\RsqL (\tchipar + \tchipara - \tchiperp - \tchiperpa)
 (\bGzero \bcdot \bo) (\bo \bcdot \bHzero)
 + \RsqL (\tchiperp + \tchiperpa) \bGzero \bcdot \bHzero]. \labelp{eq:33}
\end{eqnarray}
The force moment is determined by substituting the same transformations in the first line in
equation \ref{eq:293},
\begin{eqnarray}
 \barK_{ij} & = & \tfrac{1}{4} \mu_0 \RsqL \left( \tfrac{2}{3} \textcolor{red}{(\tchipar + \tchipara-
 \tchiperp-\tchiperpa) \ho_i \ho_l \Hzero_l} \textcolor{blue}{\Hzero_j} \right. \nonumber \\
 & & \left. \mbox{}
 - \tfrac{1}{3} \textcolor{blue}{\Hzero_i}
 \textcolor{red}{(\tchipar + \tchipara - \tchiperp - \tchiperpa) \ho_j \ho_l \Hzero_l} \right.
 \left. \mbox{} + \tfrac{1}{3} \delta_{ij}
 \textcolor{blue}{\Hzero_k} \textcolor{red}{(\tchipar + \tchipara - \tchiperp - \tchiperpa)
 \ho_k \ho_l \Hzero_l} \right. \nonumber \\
 & & \left. \mbox{} + \tfrac{2}{3} \textcolor{red}{(\tchiperp+\tchiperpa) \Hzero_i} \textcolor{blue}{\Hzero_j}
 - \tfrac{1}{3} \textcolor{blue}{\Hzero_i}
 \textcolor{red}{(\tchiperp + \tchiperpa) \Hzero_j}  
 + \tfrac{1}{3} \delta_{ij}
 \textcolor{blue}{\Hzero_k} \textcolor{red}{(\tchiperp + \tchiperpa)
 \Hzero_k} \right) \nonumber \\
 & = &   \tfrac{1}{4} \mu_0 \RsqL (\tchipar + \tchipara - \tchiperp - \tchiperpa) (
 \tfrac{2}{3} \ho_i \ho_l \Hzero_l \Hzero_j
 - \tfrac{1}{3} \Hzero_i
 \ho_j \ho_l \Hzero_l + \tfrac{1}{3} \delta_{ij} \ho_k \Hzero_k \ho_l \Hzero_l)
 \nonumber \\ & &
 \mbox{} + \tfrac{1}{12} \mu_0 \RsqL (\tchiperp + \tchiperpa) [\Hzero_i
 \Hzero_j + \delta_{ij} \Hzero_k \Hzero_k].  \labelp{eq:36}
\end{eqnarray}

The force moment can be separated into the symmetric ($\barKs$) and antisymmetric ($\barKa$) 
parts,
\begin{eqnarray}
 \barKs_{ij} & = &
  \tfrac{1}{24} \mu_0 \RsqL (\tchipar + \tchipara - \tchiperp - \tchiperpa) ( (\ho_i \Hzero_j + \Hzero_i \ho_j) \ho_l \Hzero_l
 + 2 \delta_{ij} \ho_k \Hzero_k \ho_l \Hzero_l)
 \nonumber \\ & &
 \mbox{} + \tfrac{1}{12} \mu_0 \RsqL (\tchiperp + \tchiperpa) (\Hzero_i \Hzero_j + \delta_{ij} \Hzero_k \Hzero_k),
 \labelp{eq:36}
\end{eqnarray}
\begin{eqnarray}
 \barKa_{ij} & = &
  \tfrac{1}{8} \mu_0 \RsqL (\tchipar + \tchipara - \tchiperp - \tchiperpa) (\ho_i \Hzero_j - \Hzero_i \ho_j) \ho_l \Hzero_l.
 \labelp{eq:37}
\end{eqnarray}
The torque on the particle is
\begin{eqnarray}
 T_i & = & \epsilon_{ijk} \barKa_{jk} = \tfrac{1}{4} \mu_0 \RsqL
 (\tchipar + \tchipara - \tchiperp - \tchiperpa) \epsilon_{ijk} \ho_j \Hzero_k \ho_l \Hzero_l.
 \labelp{eq:38}
\end{eqnarray}

The susceptibilities $\tchipar$ and $\tchiperp$ are evaluated in appendix \ref{appendix:rod}
using the procedure in \cite{landaulifshitz},
\begin{eqnarray}
 \tchiperp & = & 2 \tchipar = \mbox{} - 2 \pi \left( 1 - \frac{2 J_1(\tbe R)}{\tbe R J_0(\tbe R)} \right),
 \labelp{eq:39}
\end{eqnarray}
where $J_0$ and $J_1$ are Bessel functions of the first and second order.
The force, symmetric force moment and torque, equations \ref{eq:33}, \ref{eq:36} and \ref{eq:38}, are expressed in vector notation,
\begin{eqnarray}
\barbF & = & \mbox{} - \barGamma_r \mu_0 \RsqL (\bGzero \bcdot \bHzero - \tfrac{1}{2} (\bGzero \bcdot \bo) (\bHzero \bcdot \bo)),
\labelp{eq:310}
\end{eqnarray}
\begin{eqnarray}
 \barbKs & = & \mbox{} - \tfrac{1}{3} \barGamma_r \mu_0 \RsqL [\bHzero \bHzero + \bI (\bHzero \bcdot \bHzero) -
 \tfrac{1}{2} (\bo \bHzero + \bHzero \bo) (\bHzero \bcdot \bo) - \bI (\bo \bcdot \bHzero)^2],
 \labelp{eq:311}
\end{eqnarray}
\begin{eqnarray}
 \bT & = & \mbox{} \tfrac{1}{2} \mu_0 \RsqL \barGamma_r (\bo \btimes \bHzero) (\bo \bcdot \bHzero),
 \labelp{eq:312}
\end{eqnarray}
where
\begin{eqnarray}
 \barGamma_r & = & \mbox{} - \tfrac{1}{4} (\tchiperp + \tchiperpa) =
 \mbox{} \pi \left( 1 - \frac{J_1(\tbe)}{\tbe J_0(\tbe)} -
 \frac{J_1(\tbea)}{\tbea J_0(\tbea)} \right).
 \labelp{eq:313}
\end{eqnarray}

The oscillatory stress is determined by substituting $\tbH^\ast
\rightarrow \tbH$ in equation \ref{eq:019} to obtain \ref{eq:020}; the latter is
then used to determine the force and the force moments. This is equivalent to
the substitutions $\tchipara \rightarrow \tchipar$ and $\tchiperpa \rightarrow \tchiperp$
in equations \ref{eq:310}, \ref{eq:311} and \ref{eq:312},
\begin{eqnarray}
\tbF & = & \mbox{} - \tGamma_r \mu_0 \RsqL (\bGzero \bcdot \bHzero - \tfrac{1}{2} (\bGzero \bcdot \bo) (\bHzero \bcdot \bo)),
\labelp{eq:314}
\end{eqnarray}
\begin{eqnarray}
 \tKs & = & \mbox{} - \tfrac{1}{3} \tGamma_r \mu_0 \RsqL [\bHzero \bHzero + \bI (\bHzero \bcdot \bHzero) -
 \tfrac{1}{2} (\bo \bHzero + \bHzero \bo) (\bHzero \bcdot \bo) - \bI (\bo \bcdot \bHzero)^2],
 \labelp{eq:315}
\end{eqnarray}
\begin{eqnarray}
 \tbT & = & \mbox{} \tfrac{1}{2} \tGamma_r \mu_0 \RsqL (\bo \btimes \bHzero) (\bo \bcdot \bHzero),
 \labelp{eq:316}
\end{eqnarray}
where
\begin{eqnarray}
 \tGamma_r & = & \mbox{} - \tfrac{1}{2} \tchiperp =
 \mbox{} \pi \left( 1 - \frac{2 J_1(\tbe R)}{\tbe R J_0(\tbe R)} \right).
 \labelp{eq:317}
\end{eqnarray}

\begin{figure}
\psfrag{x}[][][1][0]{{$\beta R$}}
\psfrag{y}[][][0.8][0]{\textcolor{blue}{$\barGamma_r=\mbox{Re}(\tGamma_r)$}, \textcolor{red}{$\mbox{Im}(\tGamma_r)$}}

 \includegraphics[width=.49\textwidth]{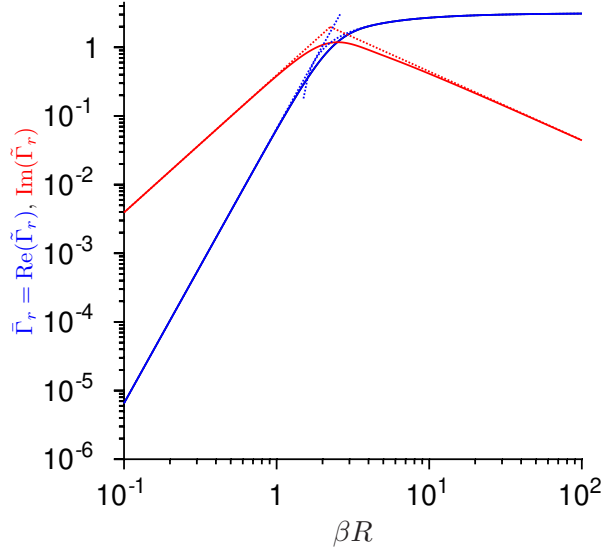}
 \caption{\label{fig:fieldgradient2} The coefficients $\barGamma_r = \mbox{Re}(\tGamma_r)$ and
 $\mbox{Im}(\tGamma_r)$ as a function of $\beta R$. The dashed blue lines on the left and right are
 the asymptotic results \ref{eq:321} and \ref{eq:322} respectively, and the dashed red lines on
 the left and right are the asymptotic results \ref{eq:323} and \ref{eq:324} respectively.}
\end{figure}

The coefficients $\barGamma_r = \mbox{Re}(\tGamma_r)$ and $\mbox{Im}(\tGamma_r)$ are shown
as a function of $\beta R$ in figure \ref{fig:fieldgradient2}. In the limits $\beta R \ll 1$
and $\beta R \gg 1$, the asymptotic expressions for $\barGamma_r$ and $\mbox{Im}(\tGamma_r)$
are as follows,
\begin{eqnarray}
 \barGamma_r = \mbox{Re}(\tGamma) & = & \frac{\pi (\beta R)^4}{48} \: \: \: \mbox{for}
 \: \: \: \beta R\ll 1,
 \labelp{eq:321}\\
 & = & \pi \left( 1 - \frac{\sqrt{2}}{(\beta R)} \right) \: \: \: \mbox{for} \: \: \: \beta R \gg 1, \labelp{eq:322} \\
 \mbox{Im}(\tGamma) & = & \frac{\pi (\beta R)^2}{8} \: \: \: \mbox{for} \: \: \: \beta R \ll 1, \labelp{eq:323} \\
 & = & \frac{\pi \sqrt{2}}{(\beta R)} \: \: \: \mbox{for} \: \: \: \beta R \gg 1. \labelp{eq:324}
\end{eqnarray}
The asymptotic results are qualitatively similar to those for a spherical particle,
\ref{eq:225}-\ref{eq:230}. Here also, it is observed that the response is in phase with the
oscillating magnetic field for $\beta R \gg 1$, while it is out of phase for $\beta R \ll 1$.
The asymptotic results \ref{eq:321} and \ref{eq:322} for $\barGamma_r$ apply to $\beta R \lesssim 1$
and $\beta R \gtrsim 3$ respectively, and the results \ref{eq:323} and \ref{eq:324} for $\mbox{Im}(\tGamma_r)$
apply to $\beta R \lesssim 1$ and $\beta R \gtrsim 10$ respectively.

\begin{figure}
 \includegraphics[width=.5\textwidth]{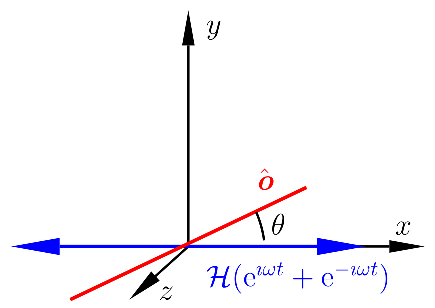}
 \caption{\label{fig:torque} Schematic for calculating the torque on a conducting
 rod oriented with orientation vector $\bo$ oriented at an angle $\theta$
 relative to the oscillating magnetic field of amplitude $\bHzero$.}
\end{figure}

The change in orientation of the rod relative to the magnetic field in a uniform magnetic field
can be inferred from equation \ref{eq:316}. It should be noted that both the orientation vector
and the oscillating field vector are apolar, that is, reversing the direction of either of the
two does not alter the force or torque. Consider the configuration where the orientation vector
$\bo$ is displaced by an angle $\theta$ in the anti-clockwise direction relative to the direction of
the applied field $\bHzero$, as shown in figure \ref{fig:torque}.
The magnetic field is in the $\bme_x$ direction, and the orientation
vector in the $x-y$ plane is $\cos{(\theta)} \bme_x + \sin{(\theta)} \bme_y$. The vector
products of the orientation vector and magnetic field are, $\bo \bcdot \bHzero = |\bHzero|
\cos{(\theta)}$, $\bo \btimes \bHzero = \mbox{} - |\bHzero| \sin{(\theta)} \bme_z$.
The torque in the $\bme_z$ direction is,
\begin{eqnarray}
 T_z & = & \tfrac{1}{2} \barGamma_r \mu_0 \RsqL |\bHzero|^2 (\mbox{} - \sin{(\theta)} \cos{(\theta)}) =
 \mbox{} - \tfrac{1}{4} \barGamma_r \mu_0 \RsqL |\bHzero|^2 \sin{(2 \theta)}. \labelp{eq:325}
\end{eqnarray}
The torque is zero for $\theta = 0$ where the orientation is parallel to the magnetic
field and $\theta = \pi/2$ where the orientation is perpendicular to the magnetic field.
The stability of these two steady orientations is determined from the change in torque
due to a small angular displacement $\Delta \theta$,
\begin{eqnarray}
 \Delta T_z & = & \mbox{} - \tfrac{1}{2} \barGamma_r \mu_0 \RsqL |\bHzero|^2 \cos{(2 \theta)} \Delta \theta.
 \labelp{eq:326}
\end{eqnarray}
The torque is in the opposite direction to $\Delta \theta$ when the perturbation
is about $\theta = 0$, acting to restore the steady orientation. The
torque is in the same direction as $\Delta \theta$ for $\theta = \pi/2$,
acting to increase the initial displacement. Thus, there is a stable steady state
when the orientation and magnetic field are parallel and an unstable steady
state when the orientation and magnetic field are perpendicular. Thus, the magnetic
torque aligns the particle in the direction of the magnetic field.

The relaxation rate of the orientation towards the magnetic field direction
is estimated from a torque balance equation for a viscous fluid. The relation
between the torque and angular velocity for a right circular cylinder with large
aspect ratio \citep{brenner}. Other shapes such as a prolate spheroid
of high aspect ratio could also be considered; only the numerical coefficients
change and the scalings do not change. The aspect
ratio of the thin rod is defined as $r = (L/2R)$, where $L$ is the length,
$L/2$ is the polar radius, and $R$ is the equatorial radius. The hydrodynamic
torque exerted by the fluid on the particle is
\begin{eqnarray}
 \bTh & = & \mbox{} - \frac{2 \eta r^2 V_p \bOmega}{9 \log{(r)}} =
 \mbox{} - \frac{\pi \eta L^3 \bOmega}{18 \log{(L/2R)}}, \labelp{eq:327}
\end{eqnarray}
where $\eta$ is the fluid viscosity, $\bOmega$ is the angular velocity
and $V_p = \pi R^2 L$ is the volume of the particle.  
In equation \ref{eq:327}, smaller terms
of $O(1/\log{(r)})$ have been neglected, and there is a negative sign because
$\bTh$ is the torque exerted by the fluid on the particle. In the viscous limit,
angular velocity is estimated from the torque balance condition,
$\barbT + \tbH = 0$,
\begin{eqnarray}
 \frac{\pi \eta L^3 |\bOmega|}{18 \log{(L/2R)}} & \sim & \tfrac{1}{2}
 \barGamma_r \mu_0 \RsqL |\tbH|^2
 \sim \tfrac{1}{2} \pi \mu_0 R^2 L |\bHzero|^2. \labelp{eq:328}
\end{eqnarray}
In the second step, we have used the estimate $\barGamma_r = \pi$ for large
$\beta R$. Therefore, the characteristic angular velocity is
\begin{eqnarray}
 |\bOmega| & \sim & \frac{9 R^2 \log{(L/2R)} \mu_0 |\bHzero|^2}{\eta L^2}.
 \labelp{eq:329}
\end{eqnarray}
Therefore, the characteristic rotational relaxation time $\tau_r$ is the inverse of the
angular velocity,
\begin{eqnarray}
 \tau_r & = & \frac{\eta L^2}{9 R^2 \log{(L/2R)} \mu_0 |\bHzero|^2}.
 \labelp{eq:330}
\end{eqnarray}


The velocity of the rod in a viscous flow is estimated from a force balance between the
magnetic and hydrodynamic drag forces. The drag force on the particle moving with
velocity $\bv$ is
\begin{eqnarray}
\bFh & = & \frac{\bo (\bo \bcdot \bv)}{\Mupar} + \frac{(\bI - \bo \bo) \bcdot \bv}{\Muperp}, \labelp{eq:332}
\end{eqnarray}
where the mobilities $\Mupar$ and $\Muperp$ are (\cite{brenner}),
\begin{eqnarray}
 \Mupar & = & \frac{\log{(L/R)}}{2 \pi L \eta}, \: \: \: \Muperp = \frac{\log{(L/R)}}{4 \pi L \eta}.
 \labelp{eq:332a}
\end{eqnarray}
The first term on the right of equation \ref{eq:322} is the drag force due to the velocity component along the
axis, and the second is the drag force due to the component perpendicular to the axis.
The mobility along the axis $\Mupar$ is twice that perpendicular to the axis $\Muperp$.
An interesting parallel here is that the drag coefficient (inverse of mobility)
for motion perpendicular to
the axis is twice that for motion parallel to the axis, just as the induced
magnetic moment for the electric field oscillation perpendicular to the axis $\tchiperp$
is twice that parallel to the axis.

The characteristic velocity is estimated by balancing the magnetic and hydrodynamic
forces. For motion along the axis, the characteristic velocity is
the product of the magnetic force along the axis in equation \ref{eq:332}
mobility $\Mupar$ (equation \ref{eq:332a}),
\begin{eqnarray}
 |\bv| & \sim & \mbox{} \frac{\barGamma_r \mu_0 \RsqL |\bHzero|^2 \log{(L/R)}}{2 \pi L L_H \eta}
 \sim \frac{R^2 \mu_0 |\bHzero|^2 \log{(L/R)}}{4 L_H \eta}. \labelp{eq:333}
\end{eqnarray}
Here, we used the estimate $\barGamma_r = \pi$ for large $\beta R$,
and $|\bG| \sim |\bHzero|/L_H$, where $L_H$ is the length scale for the variation of
the magnetic field. The characteristic time for translation over a distance
equal to the length of the rod is
\begin{eqnarray}
 \tau_t & \sim & \frac{L}{|\bv|} \sim \frac{4 L_H L \eta}{R^2 \mu_0 |\bHzero|^2
 \log{(L/R)}}.
 \labelp{eq:334}
\end{eqnarray}

Comparing the time scales \ref{eq:330} and \ref{eq:334}, it is inferred that
the time scale for translation over a distance $L$ is much larger than that
for rotation for $L_H \gg L$, that is, the length scale for magnetic field variation
is much larger than the rod length. Thus, the thin rotates and
aligns relatively quickly with the magnetic field direction, while it takes much
longer for translation over a length comparable to the length of the rod.

Due to the relatively fast rotation of the rod, it can be assumed that the rod
is aligned with the local magnetic field direction. The equation for the
force, \ref{eq:314}, reduces to
\begin{eqnarray}
 \barbF & = & \mbox{} - \tfrac{1}{2} \barGamma_r \mu_0 \RsqL \bG \bcdot \bHzero.
 \labelp{eq:335}
\end{eqnarray}
This is of the same form as equation \ref{eq:24} for a sphere.
For magnetic field variation of the form \ref{eq:42}, the force on the rod is
\begin{eqnarray}
\barbF & = & \mbox{} - \tfrac{1}{2} \barGamma_r
(\gamma_x^2 x \bme_x + \gamma_y^2 y \bme_y + \gamma_z^2 z \bme_z). \labelp{eq:336}
\end{eqnarray}
Thus, the force is directed to the origin, where the magnetic field
is minimum.

\section{Particle interactions}
\label{sec:particleinteractions}
The effect of interactions between particles
in a stationary fluid and an oscillating magnetic field is considered using a
continuum description based on the number density field $\rhon$ of the particles.
The suspension is dilute, that is, the volume fraction of the particles is small,
so pair-wise interactions between particles are included in the calculation.
The conservation equation for the number density is
\begin{eqnarray}
 \frac{\partial \rhon}{\partial t} + \bnabla \bcdot (\rhon \bv) & = &
 \DB \bnabla^2 \rhon, \labelp{eq:51}
\end{eqnarray}
where $\DB$ is the Brownian diffusivity, and $\bv$ is the velocity field
of the particles. In the base state, there is a spatially uniform number density of
particles, $\barrhon$, and the fluid velocity is zero. There is no net force
on a particle in a uniform suspension due to symmetry. However, a perturbation
of the density field $\drhon$ causes an asymmetry in particle interactions
and a force on the particles, which leads to a perturbation
of the particle velocity $\dbv$. The conservation equation \ref{eq:51}
is linearised in the perturbations,
\begin{eqnarray}
\frac{\partial \drhon}{\partial t} + \bnabla \bcdot (\barrhon \dbv) & = &
\DB \bnabla^2 \drhon. \labelp{eq:52}
\end{eqnarray}
The velocity perturbation $\dbv$ is expressed in terms of density gradients
by considering inter-particle interactions.

The conservation equation \ref{eq:52} is expressed in Fourier space using
the transform,
\begin{eqnarray}
 \star_{\bk} & = & \int \total \bx \star(\bx) \texte^{\imath \bk \bcdot \bx},
 \labelp{eq:53}
\end{eqnarray}
where the subscript $_{\bk}$ is used for the Fourier transformed quantities.
The inverse Fourier transform is
\begin{eqnarray}
 \star(\bx) & = & (2 \pi)^{-3} \int \total \bk \: \star_{\bk} \texte^{- \imath \bk
 \bcdot \bx}. \labelp{eq:54}
\end{eqnarray}
When the Fourier transform \ref{eq:53} is applied to equation \ref{eq:51}, we
obtain
\begin{eqnarray}
\frac{\partial \drhon_{\bk}}{\partial t} - \imath \bk \bcdot (\barrhon \dbv_{\bk}) & = &
\mbox{} - \DB k^2 \drhon_{\bk}. \labelp{eq:55}
\end{eqnarray}
The velocity $\dbv_{\bk}$ due to particle interactions has to be determined.

The interactions between conducting particles in an oscillating magnetic
field are of two types, the magnetic interaction due to the oscillating
dipoles and the hydrodynamic interaction due to the force moment.
In the viscous limit, the velocity field generated by a force or a force moment
is the solution of the Stokes equations. The fluid velocity at the location $\bx$
due to the force moment at the location $\bxp$ is incompressible. The particle
velocity field $\bv$, which is the ratio of the fluid velocity and the Stokes
drag coefficient. Therefore, the divergence of the particle velocity field
in equation \ref{eq:52} is also zero, and therefore there is no modulation
of number density fluctuations due to hydrodynamic interactions. Attention
is restricted to the effect of magnetic interactions.
\subsection{Spherical particles}
\label{subsec:particleinteractionssphere}
The magnetic dipole moment due to a particle at the location $\bxp$ results
in a net force on a particle with center at the location $\bx$. This
is calculated by integrating over the spherical surface $S_I$ shown in figure
\ref{fig:lengthsph}. Here, it is assumed that the radius $R_I$ of the surface
is large compared to the particle radius, but small compared to the average
separation between the particles. The perturbation to the magnetic
field at the location $\bx + \bxs$ due to a particle at the location $\bxp$ is,
\begin{eqnarray}
 \dbH_I(\bx + \bxs,\bxp) & = & \tfrac{1}{4 \pi} \bPhitwo(\bx + \bxs - \bxp) \bcdot \bHzero \times
\tfrac{1}{2} \Rcu(\tchi \texte^{\imath \omega t} + \tchia \texte^{-\imath \omega t}).
 \labelp{eq:510}
\end{eqnarray}
Here, the subscript $_I$ is used for the disturbance to the magnetic field
$\tbHp_I$ to indicate that this is due to inter-particle interactions.
When the particle separation is larger than the radius $R_I$ of the surface,
the perturbation amplitude $\tbHp$  is expressed in a Taylor series in $\bxs$,
\begin{eqnarray}
\dbH_I(\bx + \bxs, \bxp)
& = & \tfrac{1}{4 \pi} \left(\bPhitwo(\bx - \bxp) +
\bPhithree(\bx-\bxp) \bcdot \bxs \right) \bcdot \bHzero \nonumber \\
& & \times
\tfrac{1}{2} \Rcu (\tchi \texte^{\imath \omega t} + \tchia \texte^{-\imath \omega t}).
 \labelp{eq:511}
\end{eqnarray}
where $\bPhithree(\bx-\bxp) = \bnabla \bPhitwo(\bx - \bxp)$, the gradient
is with respect to $\bx$.

It should be noted that $\dbH_I(\bx + \bxs,\bxp)$ is the disturbance over the surface
of radius $R_I$ centered at $\bx$ due to another particle located at $\bxp$.
The total perturbation due to a distribution of
particles with density $\barrhon + \drhon(\bxp)$ is calculated by multiplying $\tbHp_I(\bx+ \bxs, \bxp)$
and the number density and integrating over all space,
\begin{eqnarray}
 \dbH_I(\bx + \bxs) & = & \int \total \bxp (\barrhon + \drhon(\bxp)) \dbH_I(\bx + \bxs, \bxp)
 \nonumber \\ & = &
 \int \total \bxp (\barrhon + \drhon(\bxp))
 \times \tfrac{1}{4 \pi} \left(\bPhitwo(\bx - \bxp) +
\bPhithree(\bx-\bxp) \bcdot \bxs \right) \bcdot \bHzero \nonumber \\ & &
\times \tfrac{1}{2} \Rcu (\tchi \texte^{\imath \omega t} + \tchia \texte^{-\imath \omega t}).
\labelp{eq:512}
\end{eqnarray}

There is also a perturbation to the magnetic field at the surface $\bx + \bxs$
due to the magnetisation of the medium, that is,
the modification of the background magnetic field by the distribution of
magnetic dipoles. The magnetic field in the suspension is expressed as
\begin{eqnarray}
 \bH & = & \tfrac{1}{2} \bHzero [(\texte^{\imath \omega t} + \texte^{- \imath \omega t}) +
 \Rcu (\tchi \texte^{\imath \omega t} + \tchia \texte^{- \imath \omega t})(\barrhon + \drhon(\bx + \bxs))].
\labelp{eq:513}
\end{eqnarray}
The first term in the square brackets on the right is the applied magnetic field, and
the second term is the total magnetic moment per unit volume due to the conducting
particles with number density $\barrhon + \drhon$, each particle
having moment $\tfrac{1}{2} \bHzero \Rcu (\tchi \texte^{\imath \omega t} +
\tchia \texte^{- \imath \omega t})$.
The correction to the magnetic field amplitude due to the presence of particles is
\begin{eqnarray}
 \dbH_\rho(\bx+\bxs) & = & \tfrac{1}{2} \bHzero \Rcu (\tchi \texte^{\imath \omega t} +
 \tchia \texte^{- \imath \omega t}) (\barrhon + \drhon(\bx + \bxs)).
 \labelp{eq:514}
\end{eqnarray}
Here, the subscript $_{\rho}$ in $\tbHp_\rho$ is used to indicate that the disturbance
is due to variations in the number density.
The perturbation to the number density $\drhon(\bx + \bxs)$ is expressed using a gradient
expansion in $\bxs$,
\begin{eqnarray}
 \dbH_\rho(\bx+\bxs) & = & \tfrac{1}{2} \bHzero \Rcu (\tchi \texte^{\imath \omega t} +
 \tchia \texte^{- \imath \omega t}) (\barrhon + \drhon(\bx) + \bxs \bcdot \bnabla \drhon(\bx)).
\labelp{eq:515}
 \end{eqnarray}

The total perturbation of the magnetic field, which is the sum of $\dbH_I$ and $\dbH_\rho$,
is expressed using the gradient expansion in $\bxs$,
\begin{eqnarray}
 \dbH(\bx+\bxs) & = & \dbH_I(\bx + \bxs) + \dbH_{\rho}(\bx + \bxs) \nonumber \\
 & = & \tfrac{1}{2} (\tbHp(\bx) \texte^{\imath \omega t} + \tbHpa(\bx) \texte^{- \imath \omega t})
 \nonumber \\ & & \mbox{} +
 \tfrac{1}{2} (\tbGp(\bx) \texte^{\imath \omega t} + \tbGpa(\bx) \texte^{- \imath \omega t}) \bcdot \bxs,
 \labelp{eq:516}
\end{eqnarray}
where the vector $\tbHp$ is,
\begin{eqnarray}
 \tbHp(\bx) & = & (\barrhon + \drhon(\bx)) \Rcu \tchi \bHzero+
 \int \total \bxp (\barrhon + \drhon(\bxp)) \times \tfrac{1}{4 \pi} \Rcu \tchi \bPhitwo(\bx - \bxp)
 \bcdot \bHzero,
 \labelp{eq:517}
\end{eqnarray}
and the second order tensor $\tbGp$ is
\begin{eqnarray}
 \tbGp(\bx) & = & \Rcu \tchi \bHzero \bnabla \drhon +
 \int \total \bxp (\barrhon + \drhon(\bxp)) \times \tfrac{1}{4 \pi} \Rcu \tchi \bPhithree(\bx - \bxp)
 \bcdot \bHzero, \labelp{eq:518}
\end{eqnarray}

The force on the particle at $\bx$ is
calculated by integrating the perturbation of the Maxwell stress due to the magnetic
field perturbation over the surface of the particle. In a uniform magnetic field,
the applied magnetic field gradient $\bGzero$ is zero, but there is a steady force
due to the gradient $\tbGp$ caused by the particle at $\bxp$, and the perturbation
of the magnetic field due magnetisation by other particles.
The expression \ref{eq:13} for the magnetic field is modified to include the field
due to the particle at the location $\bxp$,
 \begin{eqnarray}
  \tbH(\bx + \bxs) & = & \textcolor{blue}{\bHzero} +
  \textcolor{brown}{\tbHp(\bx)} + \textcolor{brown}{\tbGp(\bx) \bcdot \bxs}
  + \tfrac{1}{4 \pi} \Rcu \tchi \bPhitwos \bcdot (\textcolor{red}{\bHzero}
  + \textcolor{brown}{\tbHp(\bx)}) \nonumber \\ & &
  \hspace{.5in} \mbox{} + \textcolor{brown}{\tfrac{1}{4 \pi} R^5 \tlambda \bPhithrees \bddot \tbGp(\bx)},
  \labelp{eq:519}
 \end{eqnarray}
 Here, $\bxs$ is the displacement vector from the center to the surface $S_I$,
 which is different from $\bx$, the location of the center of the sphere.
 The blue terms in equation \ref{eq:519} are the applied field, the red terms are the
 modification to the magnetic field due to the presence of the sphere with center at $\bx$,
 and the brown terms include the effect of interactions with other spheres and the change
 in the magnetic field due to the magnetisation by other particles.

 The expression for the force is given by equation \ref{eq:16}. Here, we consider a 
 spherical surface
 $S_I$ with radius large compared to the particle radius $R_I$, but small compared to the distance
 between particles. Equation \ref{eq:18},
 which is the integral of the first term on the right in equation \ref{eq:16}, is modified 
 as follows,
\begin{eqnarray}
 \int_{S_I} \total S_I \tbH \tbH^\ast \bcdot \bns & = &
 \int_{S_I} \total S_I \left(
 \overset{{\tiny \textcircled{1}}}{[\textcolor{blue}{\bHzero} +
 \textcolor{brown}{\tbHp(\bx)}][\textcolor{brown}{
 \tfrac{1}{4 \pi} \tlambdaa \bPhithrees \bddot \tbGpa(\bx)}]} \right. \nonumber \\ & & \mbox{}
 \left. + \overset{{\tiny \textcircled{2}}}{
 [\textcolor{brown}{\tbGp(\bx) \bcdot \bxs}] [\tfrac{1}{4 \pi} \Rcu \tchia
 \bPhitwos \bcdot (\textcolor{red}{\bHzero} + \textcolor{brown}{\tbHpa(\bx)})]}
 \right. \nonumber \\ & & \left. \mbox{}
 + \overset{{\tiny \textcircled{3}}}{[\textcolor{brown}{\tfrac{1}{4 \pi} \tlambda
 \bPhithrees \bddot \tbGp(\bx)}] [\textcolor{blue}{\bHzero} + \textcolor{brown}{\tbHpa(\bx)}]}
  \right. \nonumber \\ & & \left. \mbox{}
 + \overset{{\tiny \textcircled{4}}}{[\tfrac{1}{4 \pi}
 \Rcu \tchi \bPhitwos \bcdot (\textcolor{red}{\bHzero} + \textcolor{brown}{\tbHp(\bx)})]
 [\textcolor{brown}{\tbGpa(\bx) \bcdot \bxs}}]
 \right) \bcdot \bns. \labelp{eq:520}
\end{eqnarray}
Here, $\bns = (\bxs/R_I)$ is the unit vector at the spherical surface in figure
\ref{fig:lengthsph}, and we have
neglected the equivalents of the cancelled terms in equation \ref{eq:18} which are
integrals of odd functions of $\bxs$. As discussed after equation \ref{eq:18}, the
terms $\textcircled{1}$ and $\textcircled{3}$ decrease proportional to $r^{-4}$,
and the surface area increases proportional to $r^2$, and therefore these terms
tend to zero for $r \sim R_I$. The terms $\textcircled{2}$ and $\textcircled{4}$
decrease proportional to $r^{-2}$, while the surface area decreases proportional
to $r^2$, and therefore these terms are finite at the surface $r \sim R_I$. Therefore,
only the terms $\textcircled{2}$ and $\textcircled{4}$ are retained in the integrals.

These terms are linearised in the perturbations, and the products of two brown
terms are neglected. The integral of the first term in the
brackets in the integrand in equation \ref{eq:16} is
\begin{eqnarray}
 \int_{S_I} \total S_I \tbH \tbH^\ast \bcdot \bns & = &
 \int_{S_I} \total S_I \left(
 \textcolor{brown}{(\tbGp(\bx) \bcdot \bxs)} \textcolor{red}{(\tfrac{1}{4 \pi} \Rcu \tchia
 \bPhitwos \bcdot \bHzero)}
 \right. \nonumber \\ & & \left. \mbox{}
 +
 \textcolor{red}{(\tfrac{1}{4 \pi} \Rcu \tchi \bPhitwos \bcdot \bHzero)}
 \textcolor{brown}{(\tbGpa(\bx) \bcdot \bxs)}
 \right) \bcdot \bns. \labelp{eq:521}
\end{eqnarray}
The right hand side of this equation is the same as that in equation \ref{eq:21}
with the transformation $\textcolor{blue}{\bGzero} \rightarrow
\textcolor{brown}{\tbGp(\bx)}$ or $\textcolor{brown}{\tbGp^\ast(\bx)}$.
Therefore, the result of the integral in
equation \ref{eq:521} is near identical to equation \ref{eq:21} with
the same transformation, taking care to substitute the
complex conjugate of $\textcolor{brown}{\tbGp(\bx)}$ where appropriate,
\begin{eqnarray}
 \int_{S_I} \total S_I \tH_i \tH_j^\ast n^{S}_j & = & R^3 (
 \tfrac{2}{3} \textcolor{brown}{\tGp_{ik}} \textcolor{red}{\tchia \Hzero_k} +
 \tfrac{1}{5} \textcolor{brown}{\tGp_{ik}^\ast} \textcolor{red}{\tchi \Hzero_k} +
 \tfrac{1}{5} \textcolor{brown}{\tGp_{ki}^\ast} \textcolor{red}{\tchi \Hzero_k}
 \nonumber \\ & & \hspace{.5in} \mbox{}
    - \tfrac{2}{15} \delta_{ij} \textcolor{brown}{\tGp_{kk}^\ast} \textcolor{red}{
    \tchi \Hzero_i}). \labelp{eq:522}
 \end{eqnarray}
The integral of the second term in square brackets in the integrand of
equation \ref{eq:16} is the complex conjugate of equation \ref{eq:522}.

The third term in square brackets in the integrand or equation \ref{eq:16} is
\begin{eqnarray}
 \int_{S_I} \total S_I \tbH \bcdot\tbH^\ast \bns & = &
 \int_{S_I} \total S_I \left(
 [\textcolor{brown}{\tbGp(\bx) \bcdot \bxs}] \bcdot \textcolor{red}{\tfrac{1}{4 \pi} \Rcu \tchia
 \bPhitwos \bcdot \bHzero)}
 \right. \nonumber \\ & & \left. \mbox{}
 +
 \textcolor{red}{\tfrac{1}{4 \pi} \Rcu \tchi \bPhitwos \bcdot \bHzero)} \bcdot
 \textcolor{brown}{(\tbGpa(\bx) \bcdot \bxs)}
 \right) \bns. \labelp{eq:523}
\end{eqnarray}
This is the same as \ref{eq:22} with the transformation $\textcolor{blue}{\bGzero} \rightarrow
\textcolor{brown}{\tbGp(\bx)}$ or $\textcolor{brown}{\tbGp^\ast(\bx)}$,
\begin{eqnarray}
 \int_{S_I} \total S_I \tH_j \tH_j^\ast n^{S}_i
& = & R^3 (\tfrac{1}{5} (\textcolor{brown}{\tGp_{ik}^\ast(\bx)} \textcolor{red}{\tchi} +
\textcolor{brown}{\tGp_{ik}(\bx)}
\textcolor{red}{\tchia}) \textcolor{red}{\Hzero_k} \nonumber \\ & & \mbox{} +
\tfrac{1}{5} (\textcolor{brown}{\tGp_{kk}^\ast(\bx)} \textcolor{red}{\tchi} +
\textcolor{brown}{\tGp_{kk}(\bx)}
\textcolor{red}{\tchia}) \textcolor{red}{\Hzero_i} \nonumber \\ & & \mbox{} -
\tfrac{2}{15} (\textcolor{brown}{\tGp_{ki}^\ast(\bx)} \textcolor{red}{\tchi} +
\textcolor{brown}{\tGp_{ki}(\bx)}
\textcolor{red}{\tchia}) \textcolor{red}{\Hzero_k}). \labelp{eq:524}
 \end{eqnarray}

The total force is obtained by substituting \ref{eq:520}, its complex conjugate,
and \ref{eq:522} into \ref{eq:16},
\begin{eqnarray}
 \barF_i & = & \tfrac{1}{4} \mu_0 \Rcu \left(
 \tfrac{2}{3} (\tchia \tGp_{ik} + \tchi \tGp_{ik}^\ast) \Hzero_k +
 \tfrac{1}{3} (\tchia \tGp_{ki} + \tchi \tGp_{ki}^\ast) \Hzero_k \right. \nonumber \\
 & & \left. \mbox{} -
  \tfrac{1}{3} (\tchia \tGp_{kk} + \tchi \tGp_{kk}^\ast) \Hzero_i
 \right).
 \labelp{eq:525}
\end{eqnarray}
Substituting the expression \ref{eq:518} for $\tGp_{ij}$, the force is
\begin{eqnarray}
 \barF_i
 & = & \tfrac{1}{4} \mu_0 \Rsi \left[ \tfrac{2}{3} \tchi \tchia \left(\Hzero_i \Hzero_k
 \frac{\partial \drhon}{\partial x_k} + \Hzero_k \Hzero_k
 \frac{\partial \drhon}{\partial x_i} \right) \right. \nonumber \\ & & \left. \mbox{}
 + \tfrac{1}{4 \pi} (2 \tchi \tchia) \int \total \bxp (\barrhon +
 \drhon(\bxp)) \Phithree_{ijk} (\bx-\bxp) \Hzero_j \Hzero_k \right].
 \labelp{eq:526}
\end{eqnarray}
Here, we have used the symmetry of $\Phithree_{ijk}$ with respect to
the interchange of any two indices, and $\Phithree_{iik} = 0$.
Equation \ref{eq:526} is expressed in vector notation as
\begin{eqnarray}
 \barbF
 & = & \tfrac{1}{2} \mu_0 \Rsi \tchi \tchia \left[ \tfrac{1}{3} \bHzero \bHzero \bcdot
 \bnabla \drhon \right. \nonumber \\ & & \mbox{} \left. + \tfrac{1}{3} (\bHzero \bcdot \bHzero) \bnabla \drhon
 + \tfrac{1}{4 \pi} \int \total \bxp (\barrhon +
 \drhon(\bxp)) \bPhithree(\bx-\bxp) \bcdot \bHzero \bHzero \right].
 \labelp{eq:527}
\end{eqnarray}

The Fourier transform of the steady force is calculated using equation \ref{eq:53},
\begin{eqnarray}
 \barbF_{\bk} & = & \tfrac{1}{2} \mu_0 \Rsi \tchi \tchia \left[\mbox{} -
 \tfrac{1}{3} \bHzero (\imath \bk \drhon_{\bk} \bcdot \bHzero)
  - \tfrac{1}{3} \imath \bk \drhon_{\bk} (\bHzero \bcdot \bHzero)
 + \tfrac{1}{4 \pi} \bPhithree_{\bk} \bddot \bHzero \bHzero
 \drhon_{\bk} \right],
 \labelp{eq:528}
\end{eqnarray}
where $\bPhithree_{\bk}$ is the Fourier transform of $\bPhithree(\bx)$. The
second term on the right of equation \ref{eq:517} is a convolution integral
of $\bPhithree(\bx-\bxp)$ and $\drhon(\bxp)$, and therefore the product
rule has been used for the Fourier transform.

The spherical harmonic solutions are derived in equation \ref{eq:15}. The Fourier
transform of the fundamental solution $\Phizero$ is,
\begin{eqnarray}
 \Phizero_{\bk} & = & \frac{4 \pi}{k^2}. \label{eq:529}
 \end{eqnarray}
 The harmonic $\bPhithree$ is obtained by taking the gradient or $\Phizero$ three
 times. Since the gradient of a function transforms to $- \imath \bk$ times the
 Fourier transform of the function, we obtain,
 \begin{eqnarray}
 \bPhithree_{\bk} & = & ( \mbox{} - \imath \bk) (\mbox{} - \imath \bk) (\mbox{}
 - \imath \bk) \Phizero_{\bk} = \mbox{} \frac{4 \pi \imath \bk \bk \bk}{k^2}. \label{eq:530}
\end{eqnarray}
This is substituted in equation \ref{eq:528} to obtain
\begin{eqnarray}
 \barbF_{\bk} & = &
 \tfrac{1}{2} \mu_0 \Rsi \tchi \tchia \left[\mbox{} -
 \tfrac{1}{3} \bHzero (\imath \bk \drhon_{\bk} \bcdot \bHzero)
  - \tfrac{1}{3} \imath \bk \drhon_{\bk} (\bHzero \bcdot \bHzero)
  + \frac{\imath \bk \bk \bk \bddot (\bHzero \bHzero) \drhon_{\bk}}{k^2}
  \right]. \labelp{eq:531}
\end{eqnarray}

The Fourier transform of the particle velocity due to the magnetic field disturbance is evaluated using Stokes law,
\begin{eqnarray}
  \dbv_{\bk} & = & \frac{\mu_0 \Rsi \tchi \tchia}{12 \pi \eta R}
  \left[\mbox{} -
 \tfrac{1}{3} \bHzero (\imath \bk \drhon_{\bk} \bcdot \bHzero)
  - \tfrac{1}{3} \imath \bk \drhon_{\bk} (\bHzero \bcdot \bHzero)
  + \frac{\imath \bk \bk \bk \bddot (\bHzero \bHzero) \drhon_{\bk}}{k^2}
  \right], \labelp{eq:532}
\end{eqnarray}
where $\eta$ is the fluid viscosity. Equation \ref{eq:532} for the velocity is
substituted into the mass conservation equation \ref{eq:55}, to obtain,
\begin{eqnarray}
 \frac{\partial \drhon_{\bk}}{\partial t} + \bDM \bddot \bk \bk \drhon_{\bk}  & = &
\mbox{} - \DB k^2 \drhon_{\bk}, \labelp{eq:533}
\end{eqnarray}
where the diffusion tensor due to magnetic interactions $\bDM$ is,
\begin{eqnarray}
 \bDM & = & \frac{\mu_0 \Rfi \tchi \tchia \barrhon}{12 \pi \eta}
 \left( \tfrac{2}{3} \bHzero \bHzero - \tfrac{1}{3} \bI \bHzero \bcdot \bHzero
 \right) \nonumber \\
 & = & \frac{\mu_0 \Rfi \tchi \tchia \barrhon |\bHzero|^2}{36 \pi \eta}
 \left[ \hbHzero \hbHzero - (\bI - \hbHzero \hbHzero) \right]. \labelp{eq:534}
\end{eqnarray}
Here, we have substituted $\bHzero = |\bHzero| \hbHzero$, where $\hbHzero$ is
the unit vector in the magnetic field direction.
The expression \ref{eq:534} for the diffusion tensor consists of two components,
one proportional to $\hbHzero \hbHzero$ in the direction of the magnetic field, and the
second proportional to $\bI - \hbHzero \hbHzero$ in the directions perpendicular
to the magnetic field. The diffusion coefficient in the direction of the magnetic field
is positive and, therefore, number density fluctuations in this direction decrease with time. In contrast,
the diffusion coefficient in the direction perpendicular to the magnetic field is
negative, and therefore these are unstable directions where the number density
fluctuations increase with time. This indicates that magnetic interactions result in the
amplification of disturbances in the direction perpendicular to the magnetic field,
and damping of disturbances along the magnetic field.

\begin{figure}
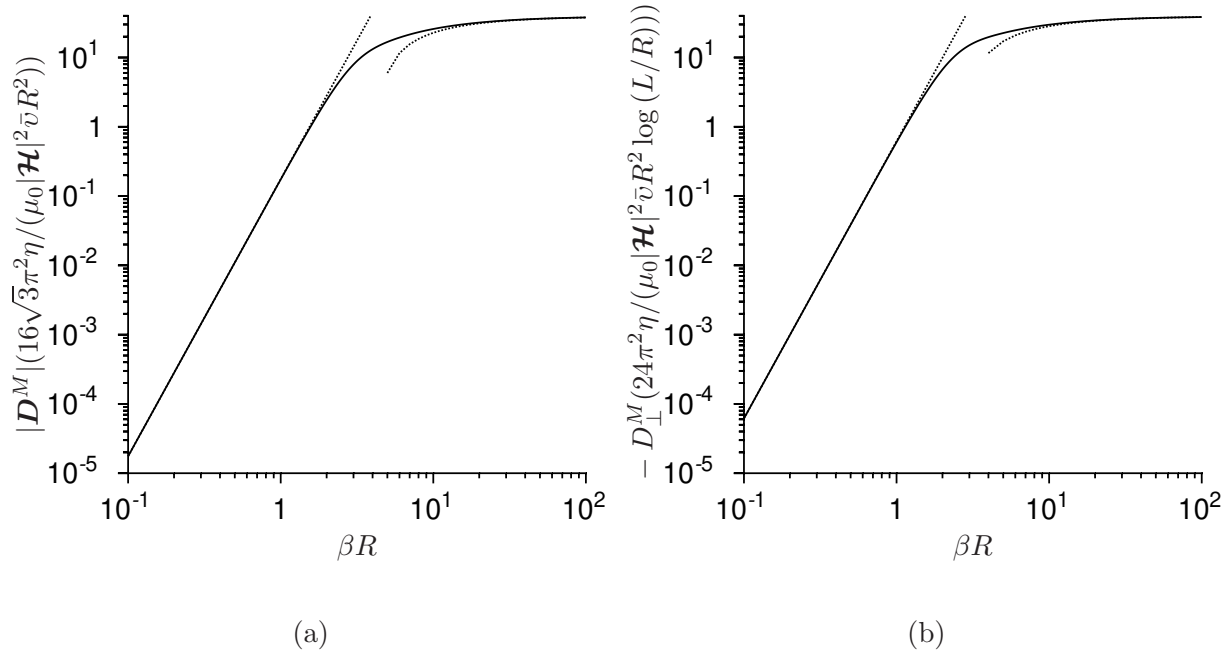

\parbox{.49\textwidth}{
\psfrag{x}[][][1][0]{{$\beta R$}}
\psfrag{y}[][][1][0]{$|\bDM| (16 \sqrt{3} \pi^2 \eta/(\mu_0
|\bHzero|^2 \barphin R^2))$}

 \includegraphics[width=.49\textwidth]{diffs.ps}

 \begin{center} (a) \end{center}
 }
\parbox{.49\textwidth}{
\psfrag{x}[][][1][0]{{$\beta R$}}
\psfrag{y}[][][1][0]{$\mbox{} - \DMperp (24 \pi^2 \eta/(\mu_0
|\bHzero|^2 \barphin R^2 \log{(L/R)}))$}

 \includegraphics[width=.49\textwidth]{diffr.ps}

 \begin{center} (b) \end{center}
 }

 \caption{\label{fig:diff} The scaled magnitude of the diffusion coefficient
 for (a) a spherical particle, $|\bDM| (16 \sqrt{3} \pi^2 \eta/\mu_0
|\bHzero|^2 \barphin R^2)$, and (b) a thin rod,
$\mbox{} - \DMperp (24 \pi^2 \eta/(\mu_0
|\bHzero|^2 \barphin R^2 \log{(L/R)}))$,
as a function of $\beta R$. In (a), the dashed lines on the
left and right are the asymptotic results \ref{eq:536} and \ref{eq:537}.
In (b), the dashed lines on the
left and right are the asymptotic result \ref{eq:564} and \ref{eq:565}.
 }
\end{figure}

The magnitude of the diffusion coefficient $\bDM$ is better understood by explicitly
specifying the $R$ dependence of the terms in equation \ref{eq:534}. The number density of
the particles is expressed as $\barrhon = \barphin/(4 \pi R^3/3)$, where
$\barphin$ is the volume fraction of the particles. With this substitution,
the magnitude of the diffusion coefficient is
\begin{eqnarray}
 |\bDM| & = & \sqrt{\bDM \bddot \bDM} =
 \frac{\mu_0 |\bHzero^2| \tchi \tchia \barphin R^2}{16 \sqrt{3} \pi^2 \eta}.
 \labelp{eq:535}
\end{eqnarray}
Figure \ref{fig:diff} (a) shows the dimensionless quantity $|\bDM| (16 \sqrt{3} \pi^2 \eta/\mu_0
|\bHzero^2| \barphin R^2)$ as a function of the parameter $\beta R$ which is the
dimensionless ratio of the particle radius and the penetration depth of the magnetic
field. In the limits $\beta R \ll 1$ and $\beta R \gg 1$, the scaled diffusion coefficient
has the form
\begin{eqnarray}
 \frac{|\bDM| (16 \sqrt{3} \pi^2 \eta)}{\mu_0 |\bHzero|^2 \barphin R^2} & \approx &
 \frac{4 \pi^2 (\beta R)^4}{225} \: \: \: \mbox{for} \: \: \: \beta R \ll 1, \labelp{eq:536} \\
 & \approx & 4 \pi^2 \left( 1 - \frac{3 \sqrt{2}}{\beta R} \right) \: \: \: \mbox{for}
 \: \: \: \beta R \gg 1. \labelp{eq:537}
\end{eqnarray}

\subsection{Thin rod}
\label{subsec:particleinteractionsrod}
In section \ref{sec:rod}, it was shown that a thin rod subject to an
oscillating magnetic field aligns with the axis in the direction of the magnetic field.
Here, we consider the force exerted as a result of particle interactions on rods aligned in
the direction of a spatially uniform oscillating magnetic field. The number density
of the rods $\rhon$ is expressed as the sum of a mean number density $\barrhon$ and
spatially non-uniform fluctuations $\drhon(\bx)$.
A fluctuation in the number density $\drhon$ could result in a change in the magnetic
field and an asymmetry in the magnetic interaction. In addition, the number density
fluctuation also results in a torque on the particles and consequently a change in
the orientation. This change in the orientation vector is determined by the condition
that the total torque on the particle is zero. There is a force generated due to the
orientation fluctuation, and this force is added to that due to the number density
fluctuation and the particle interaction in order to determine the total force
on a particle.

In the uniform state, the rods are aligned in the direction of the magnetic field.
The orientation vector is expressed as $\bo = \hbHzero
+ \dbo$, where $\hbHzero$ is the unit vector in the magnetic field direction,
and $\dbo$ is the fluctuation in the orientation vector due to spatial
non-uniformity in the number density. Since $\bo$ and $\hbHzero$
are unit vectors, $\hbHzero \bcdot \dbo = 0$ in the linear approximation.

The magnetic moment of a rod with orientation $\bo$ in a
magnetic field $\bHzero$ is
\begin{eqnarray}
 \mbox{Magnetic moment} & = &
 \tfrac{1}{2} \bo (\bo \bcdot \bHzero) \RsqL
 (\tchipar \texte^{\imath \omega t} + \RsqL \tchipara \texte^{- \imath
 \omega t}) \nonumber \\ & & \mbox{}
 + \tfrac{1}{2} (\bI - \bo \bo) \bcdot \bHzero \RsqL
  (\tchiperp \texte^{\imath \omega t} + \tchiperpa \texte^{- \imath
 \omega t}). \label{eq:551}
\end{eqnarray}
The orientation vector is expressed as $\bo = \hbHzero + \dbo$, and the equation
is linearised in $\dbo$ to obtain
\begin{eqnarray}
 \mbox{Magnetic moment} & = &
 \tfrac{1}{2} \bHzero \RsqL (\tchipar \texte^{\imath \omega t} + \tchipara \texte^{- \imath
 \omega t}) \nonumber \\ & & \mbox{}
 + \tfrac{1}{2} \dbo |\bHzero| \RsqL ((\tchipar - \tchiperp) \texte^{\imath \omega t} + (\tchiperpa
 - \tchiperpa) \texte^{- \imath \omega t}). \label{eq:552}
\end{eqnarray}
This expression for the magnetic moment is used for evaluating the amplitude of the
perturbation to the magnetic field due to interactions.

Equation \ref{eq:552} is substituted for $\tfrac{1}{2} \bHzero (\tchi \texte^{\imath \omega
t} + \tchia \texte^{- \imath \omega t})$ in equations \ref{eq:510}-\ref{eq:515}.
The perturbation to the magnetic field is given by equation \ref{eq:516} where,
instead of \ref{eq:517} and \ref{eq:518}, the amplitudes $\tbHp$ and $\tbGp$ are
\begin{eqnarray}
 \tbHp(\bx) & = & (\barrhon + \drhon(\bx)) \RsqL (\tchipar \bHzero + (\tchipar - \tchiperp)
 |\bHzero| \dbo(\bx)) \nonumber \\ & & \mbox{} +
 \RsqL \int \total \bxp (\barrhon + \drhon(\bxp)) \times \tfrac{1}{4 \pi} \bPhitwo(\bx - \bxp)
 \bcdot (\tchipar \bHzero \nonumber \\ & & \mbox{} \hspace{2in} + (\tchipar - \tchiperp) |\bHzero| \dbo(\bxp)),
 \labelp{eq:553}
\end{eqnarray}
\begin{eqnarray}
 \tbGp(\bx) & = &  \RsqL (\tchipar \bHzero + (\tchipar - \tchiperp) |\bHzero|
 \dbo) \bnabla \drhon + \RsqL \barrhon (\tchipar - \tchiperp) |\bHzero| \bnabla \dbo \nonumber \\
 & & \mbox{} + \RsqL
 \int \total \bxp (\barrhon + \drhon(\bxp)) \times \tfrac{1}{4 \pi} \tchi \bPhithree(\bx - \bxp)
 \bcdot (\tchipar \bHzero \nonumber \\ & & \mbox{} \hspace{2in}
 + (\tchipar - \tchiperp) |\bHzero| \dbo(\bxp)). \labelp{eq:554}
\end{eqnarray}

The perturbation to the orientation vector $\dbo(\bx)$ is caused by the density
variation $\drhon(\bx)$. By symmetry, the perturbation to the orientation vector is
zero if the density is uniform. This perturbation is calculated using a torque
balance equation, but this is not pursued here because it is easily verified that the
contribution to the force due to $\dbo$ is of higher order in gradients compared to
that due to $\drhon$. Since $\dbo(\bx)$ is perpendicular to $\hbHzero$,
the expression for the disturbance is necessarily of the form $\dbo(\bx) \propto
(\bI - \hbHzero \hbHzero) \bcdot \bnabla \drhon$. The contribution due to
the perturbation to the orientation vector in the expressions \ref{eq:553} and \ref{eq:554}
is one order higher in gradients compared to that due to $\drhon$. Therefore,
the contribution due to $\dbo$ is neglected in the calculation of the force on a
particle.

When $\dbo$ is neglected, equation \ref{eq:554} for $\tbGp$ is identical to equation
\ref{eq:518} for a spherical particle. The calculation of the force on the particle,
equations \ref{eq:519}-\ref{eq:527}, is also identical to that for a spherical particle
with the substitution $R^3 \tchi \rightarrow R^2 L \tchipar$ and $R^3 \tchia \rightarrow
R^2 L \tchipara$. Therefore, the equivalent of the Fourier transform of the force on the
particle, equation \ref{eq:528}, is
\begin{eqnarray}
 \barbF_{\bk} & = &
 \tfrac{1}{2} \mu_0 \RfoLsq \tchipar \tchipara |\bHzero|^2 \left[\mbox{} -
 \tfrac{1}{3} \hbHzero (\imath \bk \drhon_{\bk} \bcdot \hbHzero)
  - \tfrac{1}{3} \imath \bk \drhon_{\bk}
  + \frac{\imath \bk \bk \bk \bddot (\hbHzero \hbHzero) \drhon_{\bk}}{k^2}
  \right]. \labelp{eq:555}
\end{eqnarray}

In order to calculate the velocity disturbance, the force is resolved into
components parallel and perpendicular to the magnetic field direction $\hbHzero$,
\begin{eqnarray}
    \barbF
    & = & \hbHzero \barF_{\hbHzero} + \barbF_{\perp}, \labelp{eq:556}
\end{eqnarray}
where
\begin{eqnarray}
    \barF_{\hbHzero} & = & \tfrac{1}{2} \mu_0 \RfoLsq \tchipar \tchipara |\bHzero|^2 \drhon_{\bk}
    \left[ \mbox{} -
 \tfrac{2}{3} (\imath \bk \bcdot \hbHzero) 
  + \frac{\imath (\bk \bcdot \hbHzero)^3}{k^2}\right], \labelp{eq:557} \\
  \barbF_{\perp} & = & \tfrac{1}{2} \mu_0 \RfoLsq \tchipar \tchipara |\bHzero|^2 \drhon_{\bk}
    \left[ \mbox{} - \tfrac{1}{3} \imath \bk + \tfrac{1}{3} (\imath \bk \bcdot
    \hbHzero) \hbHzero + \frac{\imath \bk (\bk \bcdot \hbHzero)^2}{k^2} \right. \nonumber
    \\ & & \hspace{3in} \left. \mbox{} -
    \frac{\imath \hbHzero (\bk \bcdot \hbHzero)^3}{k^2}
    \right]. \labelp{eq:558}
\end{eqnarray}
The particle velocity is
\begin{eqnarray}
    \dbv_{\bk} & = & \Mupar \hbHzero \barF_{\hbHzero} + \Muperp \barbF_{\perp},
    \labelp{eq:559}
\end{eqnarray}
where $\Mupar$ and $\Muperp$ (equation \ref{eq:332a}) are the mobilities in the directions
parallel and perpendicular to the axis of the rod. Substituting \ref{eq:557} and \ref{eq:558}
for $\barF_{\hbHzero}$ and $\barbF_{\perp}$, the velocity is,
\begin{eqnarray}
    \dbv_{\bk} & = & \tfrac{1}{2} \mu_0 \RfoLsq\tchipar \tchipara |\bHzero|^2 \drhon_{\bk}
    \left[ \mbox{} - \tfrac{1}{3} \Muperp \imath \bk + 
    (\tfrac{1}{3} \Muperp - \tfrac{2}{3} \Mupar) \hbHzero (\imath \bk \bcdot 
    \hbHzero) \right. \nonumber \\ & & \left. \hspace{1.2in}
    + \frac{\Muperp \imath \bk (\bk \bcdot \hbHzero)^2}{k^2}
    \mbox{} + \frac{(\Mupar - \Muperp) \imath \hbHzero (\bk \bcdot 
    \hbHzero)^3}{k^2} \right]. \labelp{eq:559}
\end{eqnarray}
This is substituted in the mass conservation equation \ref{eq:55}, to obtain
\begin{eqnarray}
 \frac{\partial \drhon_{\bk}}{\partial t} +
 \tfrac{1}{2} \mu_0 \RfoLsq \tchipar \tchipara \barrhon |\bHzero|^2
 \left( (\tfrac{4}{3} \Muperp - \tfrac{2}{3} \Mupar)
 \hbHzero \hbHzero - \tfrac{1}{3} \bI \Muperp
 \right) \bddot \bk \bk \drhon_{\bk} & & \nonumber \\ \mbox{}
 + \frac{\tfrac{1}{2} \mu_0 \RfoLsq \tchipar \tchipara \barrhon |\bHzero|^2
 (\Mupar - \Muperp) (\bk \bcdot
    \hbHzero)^4 \drhon_{\bk}}{k^2} =
\mbox{} - \DB k^2 \drhon_{\bk}. & & \labelp{eq:560}
\end{eqnarray}

Equation \ref{eq:560} cannot be expressed as a diffusion equation due to
the third term on the left. However, a diffusion
equation can be obtained for concentration modulation with wave vector
parallel and perpendicular to the magnetic field. In the direction
perpendicular to the magnetic field, the third term on the left in
equation \ref{eq:560} is zero. The second term on the left is expressed
as $\DMperp \kperp^2 \drhon_{\bk}$, where $k_{\perp}$ is the wave vector
perpendicular to the magnetic field, and the diffusion coefficient $\DMperp$ is
\begin{eqnarray}
 \DMperp & = & \mbox{} - \tfrac{1}{6} \mu_0 \RfoLsq \tchipar \tchipara \barrhon |\bHzero|^2
  \Muperp. \labelp{eq:561}
\end{eqnarray}
It is evident that the diffusion equation $\DMperp$ is negative,
indicating that concentration fluctuations with modulation perpendicular to
the magnetic field are amplified.

For perturbations parallel to the magnetic field direction, equation \ref{eq:560}
can be reduced to a diffusion equation, where the sum of the second and third terms
on the left is of the form $\DMpar \kpar^2 \drhon_{\bk}$, where $\kpar$ is the
wave vector along the magnetic field direction, and
\begin{eqnarray}
 \DMpar & = & \tfrac{1}{3} \mu_0 \RfoLsq \tchipar \tchipara \barrhon |\bHzero|^2
  \Muperp. \labelp{eq:562}
\end{eqnarray}
Here, we have used the relation $\Mupar = 2 \Muperp$ for a thin rod.
The diffusion coefficient is positive, indicating that concentration fluctuations
with modulation along the magnetic field are damped. We also find
$\DMpar = - 2 \DMperp$ for the thin rod.

%

The dependence of the diffusivity on the length and radius of the
rod is estimated as follows.
The mobility $\Muperp$ (equation \ref{eq:332a}) is proportional to $(\log{(L/R)}/L)$,
and the susceptibility $\tchipar$ (equation \ref{eq:39}) is proportional to $R^2 L$.
The number density is expressed in terms of the volume fraction,
$\barrhon = \barphin/(\pi R^2 L)$. The magnitude of the diffusivity is
\begin{eqnarray}
 \DMperp & = & \mbox{} - \frac{\mu_0 |\bHzero|^2 \tchipar \tchipara \barphin
 R^2 \log{(L/R)}}{24 \pi^2 \eta}. \label{eq:563}
\end{eqnarray}

Figure \ref{fig:diff} (b) shows the dimensionless quantity $\DMperp (24 \pi^2 \eta/\mu_0
|\bHzero^2| \barphin R^2 \log{(L/R)})$ as a function of the parameter $\beta R$ which is the
dimensionless ratio of the particle radius and the penetration depth of the magnetic
field. In the limits $\beta R \ll 1$ and $\beta R \gg 1$, the magnitude of the scaled
diffusion coefficient has the form,
\begin{eqnarray}
 \mbox{} - \frac{\DMperp (24 \pi^2 \eta)}{\mu_0 |\bHzero|^2 \barphin R^2 \log{(L/R)}} & \approx &
 \frac{\pi^2 (\beta R)^4}{16} \: \: \: \mbox{for} \: \: \: \beta R \ll 1, \label{eq:564} \\
 & \approx & 4 \pi^2 \left( 1 - \frac{2 \sqrt{2}}{\beta R} \right) \: \: \: \mbox{for}
 \: \: \: \beta R \gg 1. \label{eq:565}
\end{eqnarray}

\section{Conclusions}
\label{sec:conclusions}
The important conclusions of this study are as follows.
\begin{enumerate} \item
There is a steady force on an electrically conducting spherical particle of radius $R$ in a spatially
varying and oscillating applied magnetic field of amplitude $\bHzero + \bGzero
\bcdot \bx$ and frequency $\omega$, where $\bx$ is the position vector from
the center of the particle.
The steady magnetophoretic force on the particle is of the form $\barF =
- \barGamma_s \mu_0 R^3 \tbH \bcdot \bGzero$, where the positive dimensionless coefficient
$\barGamma$ is a function of $\beta R = \sqrt{\mu_0 \kappa \omega R^2}$, the ratio of
the particle radius and the penetration depth of the magnetic field.
The magnetophoretic force is in the direction of decreasing magnetic field
amplitude, resulting in particle motion towards locations where the gradient of the
magnetic field is zero. This is opposite to the phenomenon of positive magnetophoresis
of magnetic particles, where the force is directed towards increasing magnetic field.

In a viscous flow, Stokes law is used to relate the velocity and the force of the
particle, and the magnetophoretic velocity of a spherical particle is proportional
to $- \tfrac{1}{2} \barGamma_s \mu_0 R^2 |\bHzero|^2/(6 \pi \eta L_H)$,
where $L_H$ is the length scale for the variation of the magnetic field.
\item
For a thin rod with radius $R$ length $L$, high aspect ratio $L \gg R$ and
orientation vector $\bo$, the magnetophoretic force is $\barbF = \mbox{} - \barGamma_r
\mu_0 R^2 L (\bGzero \bcdot \bHzero - \tfrac{1}{2} (\bGzero \bcdot \bo) (\bHzero
\bcdot \bo))$.
The force is in the direction
of decreasing magnetic field components parallel and perpendicular to the orientation
vector.

In the viscous limit, the mobility of a particle (ratio of velocity and force)
along the orientation vector is twice that perpendicular to the orientation
vector, and both are proportional to $(\eta L/\log{(L/R)})$
(\cite{brenner}). Therefore,
the induced velocity is proportional to $\mu_0 R^2 \log{(L/R)} |\bHzero|^2 / (\eta
L_H)$, where $L_H$ is the length scale for the magnetic field variation.
Thus, the appropriate length scale for magnetophoresis is
the radius of the rod, with a logarithmic correction proportional to $\log{(L/R)}$.
\item
There is a torque on a thin rod $\bT = \tfrac{1}{2} \mu_0 R^2 L \barGamma_r (\bo \btimes \bH)
(\bo \bcdot \bH)$, which tends to align the rod in the direction
parallel to the magnetic field. In a viscous flow, the induced angular velocity
is proportional to $\mu_0 |\bHzero|^2 R^2 \barGamma_r \log{(L/R)}/\eta L^2$.

The time scale for the alignment of the thin rod is
compared to the translation time over a distance comparable to the length. Here,
the mobility coefficients for a thin rod in viscous flow (\cite{brenner}) are used.
The rotation time is found to be $(L/L_H)$ smaller than the translation time 
over a distance $L$, where $L_H$ is the length scale for the magnetic
field variation. Thus, there is relatively fast orientation of the rod in the
magnetic field direction and slower magnetophoresis along the direction of
decreasing magnitude of the magnetic field.

\item The effect of far-field magnetic particle interactions
and the modification of the applied magnetic field due to particle magnetisation
is considered. For spherical particles, when there is a small spatial variation in the
particle number density, the effect of interactions reduces to an anisotropic
diffusion term in the conservation equation \ref{eq:533} for the number density.
The diffusion coefficient in the direction of the magnetic field is positive,
indicating damping of number density variations, whereas the diffusion coefficient
perpendicular to the magnetic field is positive, indicating amplification of
number density variations. This is similar to the effect of interactions in
suspensions of magnetic particles in a steady magnetic field studied in
\cite{vk22}, where it was also shown that the effect of interactions can be
reduced to an anisotropic diffusion term in the number density equation.

The components of the diffusion tensor \ref{eq:535} are proportional
to $(\mu_0 |\bHzero|^2 \barphin R^2 / \eta)$ for spherical particles,
where $\barphin$ is the volume fraction. Thus, the magnitude of the diffusion
tensor increases linearly with the volume fraction and quadratically with
particle size.
\item For a suspension of thin rods, the effect of interactions can not be reduced
to an anisotropic diffusion term in the conservation equation \ref{eq:560} for the number
density. However, in this case, it is shown that number density variations are damped along the
magnetic field and amplified perpendicular to the magnetic field.

The diffusion coefficients are proportional to $(\mu_0 |\bHzero|^2 \barphin R^2
\log{(L/R)} / \eta)$. Thus, the microscopic length scale for diffusion is the
particle radius, with a logarithmic correction proportional to $\log{(L/R)}$.

\item There are also hydrodynamic interactions between the particle, because the
Maxwell stress generates a force moment for each particle (equations \ref{eq:294},
\ref{eq:315} and \ref{eq:316}). These produce velocity fields that influence
the dynamics of neighbouring particles. However, the convective term in the number
density conservation equation, \ref{eq:52}, is zero because the
velocity field $\dbv$ obtained by solving Stokes equations has zero divergence.
Though complex phenomena such as superdiffusivity and long range flows have been
reported in anisotropic suspensions of active particles (\cite{sriram}), these
require material anisotropy in the constitutive relations for the dependence of
the flux and stress on the concentration and velocity fields.
\end{enumerate}

The coefficients $\barGamma_s, \tGamma_s \barGamma_r, \tGamma_r$ and
the components of the diffusion tensor $\bDM$
increase proportional to $(\beta R)^4$ for $\beta R \ll 1$, and they asymptote
to constants in the limit $\beta R \gg 1$. The parameter $\beta R = \sqrt{\mu_0
\kappa \omega} R$ is the ratio of the sphere/cylinder radius and the penetration
depth of the magnetic field into the particle. The magnetic permeability of
free space is $\mu_0 = 4 \pi \times 10^{-7} \mbox{kg m s}^{-2} \mbox{ A}^{-2}$.
The electrical conductivity of metals such as copper or silver is of the order
of $7 \times 10^7 \mbox{kg}^{-1} \mbox{m}^{-3} \mbox{s}^3 \mbox{A}^2$. The inverse
of the penetration depth $(\mu_0 \kappa \omega)^{-1/2}$ is $(10^2 \omega)^{-1/2}
\mbox{m}$, where $\omega$ is the angular frequency in radians per second. For
these parameter values, the length scale $\beta^{-1}$
is approximately $4$ mm when the frequency is $10^2 \mbox{Hz}$
(corresponding to the frequency of power supplies) and
approximately $400$ $\mu$m when the frequency is $10^4$ $\mbox{Hz}$
(corresponding to the frequency of acoustic waves). Thus, the
parameter $\beta R$ is $O(1)$ for particles of diameter $4$ mm for frequency
$10^2$ $\mbox{Hz}$. However, for particles of diameter $400$ $\mu$m, $\beta R$
is $O(1)$ for a much higher frequency of about $10^4$ $\mbox{Hz}$.

A convenient reference for the magnetophoretic force is the weight of the particle,
$(4 \pi R^3 \rho_m g/3)$, where $\rho_m$ is the mass density and $g$ is the gravitational
acceleration. The ratio of the magnetophoretic and gravitational forces is equal
to the ratio of the magnetophoretic velocity and the terminal velocity in a
viscous fluid. Both the weight and the
magnetophoretic force increase proportional to $R^3$, and the ratio is independent of
$R$. The magnetophoretic force and the weight ratio is $(3 \mu_0 H_0^2 / 4 \rho_m L_H)$.
It is convenient to express the ration in terms of the magnetic flux density $B_0 =
(H_0/\mu_0)$. When expressed in these units, the ratio of the magnetophoretic force
and the weight is $(3 B_0^2 / 4 \mu_0 \rho_m g L_H)$.

The ratio $(3 B_0^2 / 4 \mu_0 \rho_m g L_H)$ is shown for different parameter values
in table \ref{tab:table1}. The densities considered are $10^{3}-10^4 \mbox{kg/m}^3$
for metallic particles. The characteristic length
for the variation of the magnetic field is considered in the range $1 \mbox{cm}$
to $10 \mbox{cm}$. The acceleration due to gravity is $10 \mbox{m/s}$ and the magnetic
permeability $\mu_0 = 4 \pi \times 10^{-7} \mbox{kg m s}^{-2} \mbox{A}^{-2}$.
Table \ref{tab:table1} shows that the magnetophoretic and gravitational force
are comparable when the magnetic flux density is in the range $0.01-0.1 \mbox{T}$ for
relatively low density $\rho_m \sim 10^{3} \mbox{kg/m}^3$ and large separation $10 \,
\mbox{cm}$ or for relatively high density $\rho_m \sim 10^4 \mbox{kg/m}^3$ and
small separation $1 \, \mbox{cm}$. Thus, the magnetophoretic force is comparable
to the weight of the particle for physically realisable values of the magnetic
field and its gradient.
\begin{table}
 \begin{tabular}{|r|r|r|r|} \hline
 $L_H$ (m) & $B_0$ (T) & \multicolumn{2}{|c|}{$(3 \mu_0 H_0^2/4 \rho_m g L_H)$} \\ \hline
 & & $\rho_m = 10^3 \mbox{kg/m}^3$ & $\rho_m = 10^4 \mbox{kg/m}^3$ \\ \hline
 $10^{-1}$ & $10^{-3}$ & $6 \times 10^{-4}$ & $6 \times 10^{-5}$ \\ \hline
 $10^{-1}$ & $10^{-2}$ & $6 \times 10^{-2}$ & $6 \times 10^{-3}$ \\ \hline
 $10^{-1}$ & $10^{-1}$ & $6$ & $6 \times 10^{-1}$ \\ \hline
 $10^{-2}$ & $10^{-3}$ & $6 \times 10^{-3}$ & $6 \times 10^{-4}$ \\ \hline
 $10^{-2}$ & $10^{-2}$ & $6 \times 10^{-1}$ & $6 \times 10^{-2}$ \\ \hline
 $10^{-2}$ & $10^{-1}$ & $6 \times  10^1$ & $6$ \\ \hline
 \end{tabular}
\caption{\label{tab:table1} The ratio of the magnetophoretic force magnitude
and the weight of the particle for different values of the magnetic flux intensity
$B_0$ and the length scale $L_H$ for the variation of the magnetic field.}
\end{table}

The relative magnitude of the magnetic and Brownian diffusion for spherical
particles can be estimated from the ratio $|\bDM|/\DB$,
\begin{eqnarray}
 \frac{|\bDM|}{\DB} & = &  \frac{3 \mu_0 |\bHzero^2| \tchi \tchia \barphin R^3}{8 \sqrt{3}
 \pi k_B T}.
\end{eqnarray}
Here, $|\bDM|$ is defined in equation \ref{eq:535}, the Brownian
diffusion coefficient is $\DB = (k_B T/6 \pi \eta R)$, $k_B$ is the
Boltzmann constant and $T$ is the absolute temperature. The ratio of diffusion
coefficients is proportional to $R^3$ and is independent of the fluid
viscosity. Table \ref{tab:table2} shows the ratio of diffusivities for different
values of the magnetic flux density and particle radius. The ratio of diffusivities
changes over several orders of magnitude for particle radius in the range $10^{-5}-
10^{-3}$ m, because it is proportional to $R^7 B_0^2 \omega^2$ for $\beta R \ll 1$.
For particle size of the order of $100$ $\mu$m, the magnetic diffusion coefficient
is comparable to the Brownian diffusion coefficient even for a small magnetic flux
density $10^{-3}$ T, and relatively small frequency of $10^2$ Hz. For relatively
large magnetic flux density of $10^{-1}$ T or relatively high frequency of
$10^4$ Hz, the magnetic diffusion coefficient is much larger than the
Brownian diffusion coefficient for particle size $100$ $\mu$m.
\begin{table}
 \begin{tabular}{|r|r|r|r|r|r|} \hline
 $B_0$ (T) & $R$ (m) & $(|\bDM|/\DB)$ & $(|\bDM|/\DB)$ & $R^2/|\bDM|$ (s) & $R^2/|\bDM|$ (s)  \\
 & & $f = 10^2$ Hz & $f = 10^4$ Hz & $f = 10^2$ Hz & $f = 10^4$ Hz \\ \hline
 $10^{-3}$ & $10^{-5}$ & $7.1 \times 10^{-8} \barphin$ & $7.1 \times 10^{-4} \barphin$ &
 $6.4 \times 10^{10}/\barphin$ & $6.4 \times 10^{6}/\barphin$ \\ \hline
 $10^{-3}$ & $10^{-4}$ & $7.1 \times 10^{-1} \barphin$ & $7.1 \times 10^{3} \barphin$ &
 $6.4 \times 10^{6}/\barphin$ & $6.4 \times 10^{2}/\barphin$ \\ \hline
 $10^{-3}$ & $10^{-3}$ & $7.1 \times 10^{6} \barphin$ & $5.4 \times 10^{10} \barphin$ &
 $6.4 \times 10^2/\barphin$ & $6.4 \times 10^{-2}/\barphin$ \\ \hline
 $10^{-1}$ & $10^{-5}$ & $7.1 \times 10^{-4} \barphin$ & $7.1 \times 10^{0} \barphin$ &
 $6.4 \times 10^6/\barphin$ & $6.4 \times 10^2/\barphin$ \\ \hline
 $10^{-1}$ & $10^{-4}$ & $7.1 \times 10^{3} \barphin$ & $7.1 \times 10^{7} \barphin$ &
 $6.4 \times 10^2/\barphin$ & $6.4 \times 10^{-2}/\barphin$ \\ \hline
 $10^{-1}$ & $10^{-3}$ & $7.1 \times 10^{10} \barphin$ & $5.4 \times 10^{14} \barphin$ &
 $6.4 \times 10^{-2}/\barphin$ & $6.4 \times 10^{-6}/\barphin$ \\ \hline
 \end{tabular}
\caption{\label{tab:table2} The ratio of the magnitude of the magnetic and Brownian
and magnetic diffusion coefficient $(|\bDM|/\DB)$ and the time for diffusion over a distance
comparable to particle radius $R^2/|\bDM|$ for different values of the magnetic flux density
$B_0$ and $R$ is the particle radius. The frequency $\omega$ in rad/s is $2 \pi f$,
and the other parameters are $\mu_0 = 4 \pi
\times 10^{-7} \mbox{kg m s}^{-2} \mbox{ A}^{-2}$, $
\kappa = 7 \times 10^7 \mbox{kg}^{-1} \mbox{m}^{-3} \mbox{s}^3 \mbox{A}^2$
Boltzmann constant $k_B = 1.38 \times 10^{-23} J/K$, absolute temperature $T = 300 K$.
For the diffusion time $R^2/|\bDM|$, the assumed viscosity is $\eta = 10^{-3} \:
\mbox{kg/m/s}$.
}
\end{table}

The diffusion time, $R^2/|\bDM|$, the time taken for the particle to diffuse over
a distance comparable to its radius, is also shown in table \ref{tab:table2}. The
diffusion time also increases over several orders of magnitude for particles in the
range $10^{-5}-10^{-3}$ m, since the diffusion time scales as $R^{-4}$. The diffusion
time is less than a second for frequency of the order of $10^4$ Hz and for particle
size $100$ $\mu$m if the magnetic flux density is $10^{-1}$ T, and size above
$1$ mm if the flux density is $10^{-3}$ T. These estimates indicate that,
it is feasible to observe the anisotropic clustering in experiments if the particle
size is $100$ $\mu$m or more for high frequency magnetic fields in the range
$10^2-10^4$ Hz.

\section*{Conflicts of interest}
There are no conflicts to declare.

\section*{Acknowledgements}
This work was supported by funding from the ANRF and Ministry of Education, Government of India (Grant no. SR/S2/JCB-31/2006 and ANRF/ARG/2025/001292/ENS). The author would like to thank Prof. S. Ramaswamy for instructive discussions.

\noindent Author ORCID\\
V. Kumaran, https://orcid.org/0000-0001-9793-6523 \\

\appendix
\section{Dipole moment of a conducting sphere}
\label{appendix:sphere}
The fundamental solution for the Helmholtz equation,
\begin{eqnarray}
 \bnabla^2 \zetazero + \tbe^2 \zetazero = 0, \labelp{eq:a1}
\end{eqnarray}
which is finite at the origin $r=0$, is,
\begin{eqnarray}
 \zetazero = \frac{R \sin{(\tbe r)}}{\sin{(\tbe R)} r}. \labelp{eq:a2}
\end{eqnarray}
Here, $\zetazero$ is normalised so that $\zetazero = 1$ at $r=1$. The vector harmonic
solutions are,
\begin{eqnarray}
 \bzetaone & = & \bnabla \zetazero, \: \: \bzetatwo = \bnabla \bnabla \zetazero, ...
 \labelp{eq:a3}
\end{eqnarray}
The $n^{th}$ spherical harmonic solutions is an $n^{th}$ order tensor, which is
obtained by the action of $n$ gradients on the fundamental solution. These are
evaluated using indicial notation,
\begin{eqnarray}
\zetaone_i & = & \frac{\partial \zetazero}{\partial x_i} = \frac{\partial r}{\partial x_i}
\frac{\total \zetazero}{\total r} = \frac{x_i}{r} \frac{\total \zetazero}{\total r},
\labelp{eq:a4}
\end{eqnarray}
\begin{eqnarray}
\zetatwo_{ij} & = & \frac{\partial}{\partial x_j} \frac{\partial \zetazero}{\partial x_i}
= \frac{\partial}{\partial x_j} \left( \frac{x_i}{r} \frac{\total \zetazero}{\total r}
\right) \nonumber \\
& = & \left( \frac{\delta_{ij}}{r} - \frac{x_i x_j}{r^3} \right)
\frac{\total \zetazero}{\total r} + \frac{x_i x_j}{r^2} \frac{\total^2 \zetazero}{\total r^2}
\nonumber \\
& = & \left( \frac{\delta_{ij}}{r} - \frac{x_i x_j}{r^3} \right)
\frac{\total \zetazero}{\total r} + \frac{x_i x_j}{r^2} \left( \mbox{} - \frac{2}{r}
\frac{\total \zetazero}{\total r} - \tbe^2 \zetazero \right) \nonumber \\
& = & \left( \frac{\delta_{ij}}{r} - \frac{3 x_i x_j}{r^3} \right)
\frac{\total \zetazero}{\total r} - \frac{\tbe^2 x_i x_j \zetazero}{r^2}.
\labelp{eq:a5}
\end{eqnarray}

The magnetic field is expressed as the curl of a magnetic potential $\tbA$, $\tbH = \bnabla
\btimes \tbA$, so that the solenoidal condition $\bnabla \bcdot \tbH = 0$ is satisfied.
The applied magnetic field $\bHzero$ and gradient $\bGzero$ are pseudo vectors\footnote{The direction reverses
upon inversion between right- and left-handed co-ordinates} whereas the magnetic potential
$\tbA$ is a real vector\footnote{The direction does not change upon inversion between
right- and left-handed co-ordinates}. The general expression for the magnetic potential is,
\begin{eqnarray}
 \tbA & = & \bnabla \btimes \left( \Ci \bHzero \zetazero 
 \right), \labelp{eq:a6}
\end{eqnarray}
where $\Ci$ is a complex constant. The magnetic field is the curl of the potential,
\begin{eqnarray}
 \tbH & = & \bnabla \btimes \tbA = \bnabla \btimes \bnabla \btimes (\Ci \bHzero \zetazero) \nonumber \\
 & = & \Ci \bHzero \bcdot (\bnabla \bnabla \zetazero - \bI \bnabla^2 \zetazero)
  = \Ci (\bzetatwo \bcdot \bHzero - \tbe^2 \zetazero \bHzero).
 \labelp{eq:a7}
 \end{eqnarray}
 The expression for the magnetic field is simplified using equations \ref{eq:a3}-\ref{eq:a5}
 for $\zetazero$-$\bzetatwo$,
 \begin{eqnarray}
  \tbH & = & \Ci \left[ \left( \frac{\bHzero}{r} - \frac{3 \bx (\bHzero \bcdot
  \bx)}{r^3}
  \right) \frac{\total \zetazero}{\total r} + \tbe^2 \left( \bHzero - \frac{\bx (\bHzero
  \bcdot \bx)}{r^2} \right) \zetazero \right].
  \labelp{eq:a8}
 \end{eqnarray}

 The boundary condition is the continuity of magnetic field at the surface, $r=R$. Substituting
 the expressions \ref{eq:15a} and \ref{eq:a8} for the magnetic fields outside and inside
 the particle, and equating the coefficients of $\bHzero$ and $\bx (\bHzero \bcdot \bx)$,
 we obtain two equations for $\Co$ and $\Ci$,
 \begin{eqnarray}
  \bHzero & : & 1 - \frac{\tchi}{4 \pi R^3} = \Ci \left. \left( \frac{1}{R} \frac{\total \zetazero}{\total r} + \tbe^2 \zetazero \right) \right|_{r=R},
  \labelp{eq:a9}
 \end{eqnarray}
\begin{eqnarray}
\bx (\bHzero \bcdot \bx) & : & \frac{3 \tchi}{4 \pi R^5} = \Ci \left. \left( \mbox{} - \frac{3}{R^3}
\frac{\total \zetazero}{\total r}
- \frac{\tbe^2 \zetazero}{R^2} \right) \right|_{r=R}, \labelp{eq:a10}
\end{eqnarray}
The solutions of equations \ref{eq:a9}-\ref{eq:a10}, after substituting $\left. \zetazero \right|_{r=R} = 1$
are,
\begin{eqnarray}
        \tchi & = & \mbox{} - 2 \pi R^3 \left( 1 + \frac{3 \zetazerop}{R \tbe^2} \right), \: \: \:
  \Ci = \frac{3}{2 \tbe^2}, \labelp{eq:a11} 
\end{eqnarray}
where
\begin{eqnarray}
 \zetazerop & = & \left. \frac{\total \zetazero}{\total r} \right|_{r=R} = \frac{\tbe R \cot{(\tbe R)} - 1}{R}.
 \labelp{eq:a12}
\end{eqnarray}
Therefore, the amplitude of the magnetic susceptibility is,
\begin{eqnarray}
 \tchi & = & \mbox{} - 2 \pi \left( 1 - \frac{3}{(\tbe R)^2} + \frac{3 \cot{(\tbe R)}}{\tbe R} \right).
\end{eqnarray}

In order to determine the coefficient $\tlambda$ in equation \ref{eq:15a}, it is necessary to include the
higher order terms in equation \ref{eq:a4} that depend on $\bGzero$. However, the expressions for the force
and force moments, \ref{eq:24} and \ref{eq:294}, do not depend on $\tlambda$, and therefore this calculation is
not pursued.

\section{Dipole moment of thin rod}
\label{appendix:rod}
The dipole moment of a thin rod due to an applied oscillating field can be expressed as the superposition
of the dipole moments due to the components of the magnetic field parallel and perpendicular to the axis,
\begin{eqnarray}
\tbH & = & \tbHpar + \tbHperp, \labelp{eq:b1}
\end{eqnarray}
where $\bHzeropar = \bo (\bo \bcdot \bHzero)$ and $\bHzeroperp = (\bI - \bo \bo) \bcdot \bHzero$.
The magnetic susceptibility is a tensor of the form $L \tchipar \bo (\bo \bcdot \bHzero) +
L \tchiperp (\bI - \bo \bo) \bcdot \bHzero$, where $\tchipar$ is the susceptibility per unit length
along the cylinder axis, and $\tchiperp$ is the susceptibility per unit length perpendicular to the axis.
The components $\tchipar$ and $\tchiperp$ are calculated in a two-dimensional co-ordinate system in
the plane perpendicular to the axis for a thin rod for $L \gg 1$, where the length is much larger than
the radius.
\subsection{Dipole moment perpendicular to the axis}
In the plane perpendicular to the axis, two-dimensional polar harmonics are used to determine the
induced dipole moment. The cross section of the rod is a disk of unit radius in scaled co-ordinates
centered at the origin.
The $x-y$ co-ordinate system is used, where $r = \sqrt{x^2+y^2}$ is the distance from the origin.
The magnetic field outside the rod is irrotational and solenoidal, and therefore
it can be expressed as the gradient of a potential which is a linear function of the magnetic field
and the polar harmonics. The magnetic potential and field outside the rod are,
\begin{eqnarray}
  \tphi_H & = & \bHzeroperp \bcdot \bx + \frac{R^2 \tchiperp}{2 \pi} \bHzeroperp \bcdot \bPhionex, \labelp{eq:b2}
\end{eqnarray}
\begin{eqnarray}
 \tbHperp & = & \bHzeroperp + \frac{\tchiperp R^2}{2 \pi} \bPhitwox \bcdot \bHzeroperp, \labelp{eq:b3}
\end{eqnarray}
where $R^2 \tchiperp$ is the magnetic susceptibility per unit length of the rod perpendicular to
the $x-y$ plane, $\tchiperp$ is dimensionless,
$\Phizerox = \mbox{} log{(r)}$ is the fundamental solution in two dimensions, and the decaying
harmonics are,
\begin{eqnarray}
 \bPhionex & = & \mbox{} - \frac{\bx}{r^2}, \: \: \: \bPhitwox = \left( \frac{2 \bx \bx}{r^4} -
 \frac{\bI}{r^2} \right). \labelp{eq:b4}
\end{eqnarray}
The solution \ref{eq:b3} is expressed in terms of the harmonics \ref{eq:b4},
\begin{eqnarray}
 \tbHperp & = & \bHzeroperp \left( 1 - \frac{R^2 \tchiperp}{2 \pi r^2} \right) +
 \frac{\tchiperp \bx \bx \bcdot \bHzeroperp R^2}{\pi r^4}. \labelp{eq:b5}
\end{eqnarray}

The Helmholtz equation \ref{eq:a1} is used to evaluate the magnetic field in the rod. Since the
divergence of the magnetic field is zero, the magnetic field is expressed as the curl of the
real vector potential $\tbA$, which is expressed as,
\begin{eqnarray}
 \tbA & = & \bnabla \btimes (\Ciperp \bHzeroperp \xizero), \labelp{eq:b6}
\end{eqnarray}
where $\Ciperp$ is a complex constant, and $\xizero$ is the scalar solution of the Helmholtz equation
in two dimensions, $\bnabla^2 \xizero + \tbe^2 \xizero = 0$,
\begin{eqnarray}
\xizero & = & \frac{J_0(\tbe r)}{J_0(\tbe R)}. \labelp{eq:b7}
\end{eqnarray}
Here, $J_0(\tbe r)$ is the zeroeth order Bessel function which is finite at $r=0$, and $\xizero$ is
normalised to have the value $1$ at $r=R$. The magnetic field is,
\begin{eqnarray}
 \tbHperp & = & \bnabla \btimes \bnabla \btimes (\Ciperp \bHzeroperp \xizero) \nonumber \\
 & = & \bnabla (\bnabla \bcdot \Ciperp \bHzeroperp \xizero) - \bnabla^2 (\Ciperp \bHzeroperp \xizero)
 \nonumber \\
 & = & \Ciperp \bHzeroperp \bcdot \bxitwo + \tbe^2 \Ciperp \bHzeroperp \xizero, \labelp{eq:b8}
\end{eqnarray}
where the vector and tensor solutions for the Helmholtz equation in indicial notation are,
\begin{eqnarray}
 \xione_i & = & \frac{\partial \xizero}{\partial x_i} = \frac{x_i}{r}
 \frac{\total \xizero}{\total r}, \labelp{eq:b9}
\end{eqnarray}
\begin{eqnarray}
 \xitwo_{ij} & = & \left( \frac{\delta_{ij}}{r} - \frac{x_i x_j}{r^3}
 \right) \frac{\total \xizero}{\total r} + \frac{x_i x_j}{r^2} \frac{\total^2
 \xizero}{\total r^2} \nonumber \\
 & = & \left( \frac{\delta_{ij}}{r} - \frac{x_i x_j}{r^3}
 \right) \frac{\total \xizero}{\total r} + \frac{x_i x_j}{r^2}
 \left( \mbox{} - \frac{1}{r} \frac{\total \xizero}{\total r} - \tbe^2 \xizero \right)
 \nonumber \\
 & = & \left( \frac{\delta_{ij}}{r} - \frac{2 x_i x_j}{r^3}
 \right) \frac{\total \xizero}{\total r} - \frac{\tbe^2 x_i x_j \xizero}{r^2},
 \labelp{eq:b10}
\end{eqnarray}
The magnetic field \ref{eq:b8} is expressed in terms of the harmonics \ref{eq:b7} and
\ref{eq:b10},
\begin{eqnarray}
 \tbHperp & = & \Ciperp \left[ \bHzeroperp \left( \frac{1}{r} \frac{\total \xizero}{\total r}
 + \tbe^2 \xizero \right) - \bx (\bHzeroperp \bcdot \bx) \left( \frac{2}{r^3} \frac{\total \xizero}{
 \total r} + \frac{\tbe^2 \xizero}{r^2} \right) \right]. \labelp{eq:b11}
\end{eqnarray}

The constants $\tchiperp$ and $\Ciperp$ from the continuity condition for the magnetic field
at $r=R$, that is, by equating \ref{eq:b5} and \ref{eq:b11}. The coefficients of $\bHzeroperp$
and $\bx (\bx \bcdot \bHzeroperp)$ at $r=R$ are,
\begin{eqnarray}
 \bHzeroperp & : & 1 - \frac{\tchiperp}{2 \pi} = \Ciperp \left(\frac{\xizerop}{R} + \tbe^2 \xizero \right), \\
 \bx (\bx \bcdot \bHzeroperp) & : & \mbox{} \frac{\tchiperp}{\pi R^2} = \Ciperp \left( \mbox{} - \frac{2 \xizerop}{R^3}
 - \frac{\tbe^2 \xizero}{R^2} \right),
 \labelp{eq:b12}
\end{eqnarray}
where $\xizerop = \left. (\total \xizero/\total r) \right|_{r=R}$.
These are solved to obtain $\tchiperp$ and $\Ciperp$,
\begin{eqnarray}
 \tchiperp & = & \mbox{} - 2 \pi \left( 1 + \frac{2 \xizerop}{\tbe^2} \right) = - 2 \pi
 \left( 1 - \frac{2 J_1(\tbe R)}{\tbe R J_0(\tbe R)} \right),
 \: \: \: \Ciperp = \mbox{} \frac{2}{\tbe^2}. \labelp{eq:b13}
\end{eqnarray}
\subsection{Dipole moment parallel to the axis}
The component of the magnetic field $\Hzeropar$ parallel to the axis is uniform outside the conducting
cylinder. Within the cylinder, there is a variation in the magnetic field with radial position. The magnetic field
has the same form as \ref{eq:b8}, with $\bHzeroperp$ replaced by $\bHzeropar$. Since $\bHzeropar$ is
perpendicular to $\bzetatwo$, the magnetic field is expressed as,
\begin{eqnarray}
 \tbHpar & = & \bHzeropar \xizero.
\end{eqnarray}
The magnetic moment due to the current distribution within the cylinder is,
\begin{eqnarray}
 \tchipar \bHzeropar & = & \tfrac{1}{2} L \int_{S} \total S (\bx \btimes \tbJ), \labelp{eq:b21}
\end{eqnarray}
  where $S$ is the surface of a unit circle, $\bx$ is the position vector in the plane
  perpendicular to the axis, and $\tbJ$ is the current density given by equation \ref{eq:08}.
  Here, we assume that the length $L$ is much larger than the radius of the rod, and the
  current density is independent of the axial co-ordinate if end effects are neglected.
  Since the magnetic field is along the axis and the variation of the magnetic field is in the
  radial direction, the eddy current is,
\begin{eqnarray}
 \tbJ(\bx) & = & \mbox{} - \frac{\partial \tHpar}{\partial r} \bme_\phi = \mbox{} - \Hzeropar
 \frac{\total \xizero}{\total r} \bme_\phi. \labelp{eq:b22}
\end{eqnarray}
Therefore, the magnetic moment is,
\begin{eqnarray}
  \RsqL \tchipar \bHzeropar & = & \mbox{} - \frac{L}{2} \int_{S} \total S \bme_{\parallel} \Hzeropar r \frac{\total
  \xizero}{\total r} = \mbox{} - \pi R^2 L \bme_{\parallel} \Hzeropar \int_{0}^{1} r \total r \left( r
  \frac{\total \xizero}{\total r} \right), \labelp{eq:b23}
\end{eqnarray}
where $\bme_{\parallel}$ is the unit vector along the axis of the cylinder.
The integral can be evaluated analytically,
\begin{eqnarray}
  \RsqL \tchipar \bHzeropar & = & \mbox{} - \pi L R^2 \Hzeropar \bme_{\parallel} \left( 1 - \frac{2 J_1(\tbe R)}{\tbe R
  J_0(\tbe R)} \right). \labelp{eq:b24}
\end{eqnarray}
Since $\Hzeropar \bme_{\parallel} = \bHzeropar$, the susceptibility parallel to the axis is,
\begin{eqnarray}
  \tchipar & = & \mbox{} - \pi \left( 1 - \frac{2 J_1(\tbe R)}{\tbe R J_0(\tbe R)} \right). \labelp{eq:b24}
\end{eqnarray}
Thus, the susceptibility along the axis is one half of the susceptibility perpendicular to the axis.

\bibliographystyle{unsrt}
\bibliography{ref}
\end{document}